\documentclass[twocolumn,prb,showpacs,preprintnumbers,amsmath,amssymb]{revtex4}
\usepackage{color}
\usepackage{graphicx}
\usepackage{amssymb}

\makeatletter

\usepackage{epsfig}
\usepackage{graphicx}% Include figure files
\usepackage{dcolumn}% Align table columns on decimal point
\usepackage{bm}% bold math
\usepackage{graphicx,color}

\usepackage{ulem}
\DeclareMathOperator{\diag}{diag}

\makeatother
\begin{document}

\title{Site dilution of quantum spins in the honeycomb lattice}

\author{Eduardo V. Castro,$^{1,2}$ N. M. R. Peres,$^{1,3,4}$
K. S. D. Beach$^{1}$, and  Anders W. Sandvik$^{1}$}

\affiliation{$^{1}$Department of Physics, Boston University, 590 Commonwealth
Avenue, Boston, MA 02215, USA}

\affiliation{$^{2}$ CFP and Departamento de F\'{\i}sica, Faculdade de 
Ci\^encias,
Universidade do Porto, P-4169-007 Porto, Portugal}

\affiliation{$^{3}$Max-Planck-Institut f\"ur Physik komplexer Systeme,
N\"othnitzer Str.\ 38, 01187 Dresden, Germany}

\affiliation{$^{4}$Center of Physics and Departamento de F\'{\i}sica,
Universidade do Minho, P-4710-057, Braga, Portugal}

\date{\today{}}

\begin{abstract}
We discuss the effect of site dilution on both the magnetization and
the density of states of quantum spins in the honeycomb lattice, described
by the antiferromagnetic Heisenberg spin-$S$ model. Since the disorder
introduced by the dilution process breaks translational invariance,
the model has to be solved in real space. For this purpose a real-space
Bogoliubov-Valatin transformation is used. In this work we show that for
the  $S>1/2$ the system can be analyzed in terms of linear spin wave theory,
in the sense that for all dilution concentrations the assumptions of  validity
for the theory hold. For spin $S=1/2$, however, the linear spin wave
approximation breaks down. In this case, we have studied the effect of dilution
on the staggered magnetization using the Stochastic Series Expansion Monte
Carlo method. Two main results are to be stressed from the Monte Carlo
method: (i) a better value for the staggered magnetization of the
undiluted system, $m_{\rm av}(L\rightarrow\infty)=0.2677(6)$, relatively to
the only result available to date in the literature, and based on Trotter
error extrapolations; (ii) a finite value of the staggered magnetization
of the percolating cluster at the classical percolation threshold,
showing that there is no quantum critical
transition driven by dilution in the  Heisenberg model. In the solution of the
problem using linear the spin wave method we pay special attention to the
presence of zero energy modes. We show, for a finite-size system (in a
bipartite lattice), that if the two sub-lattices are evenly diluted the system
always has two zero energy modes, which play the role of Goldstone boson modes
for a diluted lattice, having no translation symmetry but supporting
long range magnetic order. We also discuss the case when the two sub-lattices
are not evenly diluted. In this case, for finite size lattices, 
the Goldstone modes are not a well defined concept, and special
care is needed in taking them into account 
in order for sensible physical results can be obtained. Using a
combination of linear spin wave analysis and the recursion method
we were able to obtain the thermodynamic limit behavior of the density of
states for both the square and the honeycomb lattices. We have used both
the staggered magnetization and the density of states to analyze neutron
scattering experiments (determining the effect of dilution on the system's
magnetic moment) and N\'eel temperature measurements on quasi-two-dimensional
honeycomb systems. Our results are in quantitative agreement with experimental
results on Mn$_{p}$Zn$_{1-p}$PS$_{3}$ (a diluted $S=5/2$ system) and on the
Ba(Ni$_{p}$Mg$_{1-p}$)$_{2}$V$_{2}$O$_{8}$ (a diluted  $S=1$ system).
Our work should stimulate further experimental research in Heisenberg
diluted-honeycomb systems.
\end{abstract}


\maketitle

%%%%%%%%%%%%%%%%%%%%%%%%%%%%%%%%%%%%%%%%%%%%%%%%%%%%%%%%%%%%%%%%%%%%%%%%%%%%%%%
%%%%%%%%%%%%%%%%%%%%%%%%%%%%%%%%%%%%%%%%%%%%%%%%%%%%%%%%%%%%%%%%%%%%%%%%%%%%%%%
%%%%%%%%%%%%%%%%%%%%%%%%%%%%%%%%%%%%%%%%%%%%%%%%%%%%%%%%%%%%%%%%%%%%%%%%%%%%%%%
\section{Introduction}

The study of dilution and its effect on the magnetic properties of 
antiferromagnetic materials is a central problem in modern condensed
matter theory.\cite{VMG+02,GSK+00,CCN01,MNC04,San02}
For the square lattice, a number of important experimental and theoretical
results have been reported.\cite{VMG+02,CCN01,MNC04,San02} For the honeycomb
lattice, there are some experimental results in the literature\cite{GSK+00}
already, but the corresponding theoretical understanding lags far behind.

Insulating antiferromagnets are possible candidates for exhibiting
quantum critical points separating ordered from disordered phases.
The quantum corrections to the staggered magnetization of diluted
antiferromagnetic insulators became an important experimental and
theoretical topic after site dilution of La$_{2}$CuO$_{4}$ by non
magnetic impurities,\cite{VMG+02,CCR+91} such as Zn or Mg. Theoretical
studies interpreting the magnetic properties of these diluted systems
have been recently performed,\cite{MNC04,CCN02,San02,CCN01} showing
a good agreement between theory and experiment. A description of
the effect of dilution on the spin flop phase of La$_{2}$CuO$_{4}$
was attempted from the point of view of a simple mean field 
theory,\cite{Peres03} with some qualitative agreement with experimental
results. In addition, the expectation of a magnetic quantum phase
transition driven by the interplay of dilution and quantum fluctuations
was shown not to occur in the antiferromagnetic Heisenberg model in a
square lattice.\cite{San02,MNC04}
In the undiluted case, on the other hand, it was shown that the
Heisenberg model itself is incapable of describing the high energy
part of the spin wave spectrum; a calculation starting from the
Hubbard model was shown to give the correct high energy behavior.
\cite{PA03,PA02,SSS02}

The key role played by dimensionality in determining the behavior
of a system of quantum magnetic moments lends special importance
to the honeycomb lattice, which has the lowest possible co-ordination
in more than one dimension (see Fig.~\ref{cap:honey}). Realizations
of insulating antiferromagnets based on this lattice have already
been achieved both with and without magnetic dilution. Recently Spremo
\emph{et al}.~\cite{Spr+05} have studied the magnetic properties of
a metal-organic antiferromagnet on an undiluted but distorted honeycomb
lattice. The authors found good agreement between the theoretical
predictions obtained within the framework of a modified spin wave approach
and the experimental results for the magnetization as a function of
uniform external field and for the uniform zero-field susceptibility.

Honeycomb layers are also found in transition-metal thiophosphates
MPS$_{3}$, where M is a first row transition metal. These compounds
are viewed as {}``perfect'' 2D magnetic systems because of the weak
van der Waals cohesion energy binding the layers. In each layer the
magnetic ions are arranged in a honeycomb lattice. Neutron diffraction
and magnetic susceptibility studies on MnPS$_{3}$, FePS$_{3}$, and
NiPS$_{3}$ antiferromagnets\cite{FBO+82,KSY83,JV92} ($S=5/2$, $S=2$,
and $S=1$, respectively) showed the existence of quite different
types of ordering among the different compounds. Whereas for FePS$_{3}$
and NiPS$_{3}$ the metal ions are coupled ferromagnetically to two
of the nearest neighbors and antiferromagnetically to the third, for
MnPS$_{3}$ all nearest neighbors interactions within a layer are
antiferromagnetic. In fact, it turns out that the simplest nearest-neighbor
antiferromagnetic Heisenberg model is a reasonable approximation for
the description of the magnetic properties in MnPS$_{3}$, although
the second- ($J_{2}$) and third-nearest-neighbor ($J_{3}$)
interactions---which are both also antiferromagnetic---are
not negligible for this compound ($J_{1}/J_{2}\sim10$ and
$J_{1}/J_{3} \sim4$).\cite{WRL+98} Substitution of magnetic
Mn$^{2+}$ ions by nonmagnetic Zn$^{2+}$ impurities showed that
long-range order (LRO) is lost at $p=0.46\pm0.03$ for
Mn$_{p}$Zn$_{1-p}$PS$_{3}$.\cite{CV96,GH98,GSK+00}
The fact that LRO is preserved for dilutions higher than the classical
percolation threshold for the honeycomb lattice, 
$p_{\text{c}}\simeq0.7$, is attributed to the significance
of $J_{2}$ and $J_{3}$ in this compound.

Recently, Rogado \emph{et al}.~\cite{RHL+02} have characterized the
magnetic properties of the $S=1$ honeycomb compound BaNi$_{2}$V$_{2}$O$_{8}$,
which can be described as a weakly anisotropic 2D Heisenberg
antiferromagnet.\cite{HvN+03} The magnetic Ni$^{2+}$ ions lie on
weakly coupled honeycomb layers, exhibiting antiferromagnetic
LRO close to 50 K. The doped compound 
Ba(Ni$_{p}$Mg$_{1-p}$)$_{2}$V$_{2}$O$_{8}$ has a fraction
$1-p$ of the honeycomb layer sites substituted by Mg$^{2+}$---a
nonmagnetic ion. Magnetic susceptibility studies showed that the
N\'{e}el temperature is substantially reduced with increasing doping in
the range $0.84\leq p\leq1$. For $p=0.84$ the onset of antiferromagnetic
LRO occurs only at $T_{N}\simeq17$ K, a $T_{N}$ reduction of almost
70\% relative to its undiluted value. It would be interesting
to know whether the suppression of antiferromagnetic LRO by
nonmagnetic impurities occurs at the classical percolation transition
$p_{\text{c}}\simeq0.7$, as predicted by our calculations (see below).

In addition to these exciting experimental results, the theoretical
result of Mucciolo \emph{et al}.\ for the square lattice,\cite{MNC04}
where the vanishing of the staggered magnetization for the $S=1/2$
systems coincides with the classical percolation transition, opened
the naive expectation that for a 2D lattice with nonfrustrating nearest
neighbor interactions and a smaller number of neighbors, magnetic
quantum-phase transitions driven by the interplay of disorder and
quantum fluctuations could occur. The honeycomb lattice is the simplest
realization of such a lattice, for its coordination number is smaller than
that of the square lattice. On the other hand, large-scale quantum Monte
Carlo simulations of the square lattice have shown that the percolating 
cluster actually has a robust long-range order,\cite{San02} in disagreement
with the spin wave calculation where this order vanishes very close
to the percolation point. Hence, spin wave theory is not reliable
at and close to the percolation point for $S=1/2$, and we expect
this break-down also for the honeycomb lattice at the percolation point.
This expectation is confirmed; we have performed quantum Monte Carlo 
simulations that show only a rather modest reduction of the sublattice 
magnetization of the percolating cluster, whereas there is no long-range 
order in spin wave theory for $S=1/2$ in this case.

Experimental $S=1/2$ antiferromagnetic
systems with honeycomb lattice structure have already been reported
by Zhou \emph{et al}.~\cite{ZD91} in the A$_{2}$CuBr$_{4}$ salt,
where A is morpholinium (C$_{4}$H$_{10}$NO). Their data is well
described by a nearest-neighbor antiferromagnetic Heisenberg model,
but with two different couplings $J_{a}$ and $J_{b}$. To the best of our
knowledge, the dilution of this system has not yet been attempted.

Motivated by the experimental results on diluted Mn$_{p}$Zn$_{1-p}$PS$_{3}$
and Ba(Ni$_{p}$Mg$_{1-p}$)$_{2}$V$_{2}$O$_{8}$ and by the possibility
of quantum phase transitions driven by the interplay of disorder and
quantum fluctuations, we study here the effect of site dilution on
the magnetic properties of the Heisenberg antiferromagnetic nearest-neighbor
model, for an arbitrary spin-$S$ value. Our study is performed both at zero
and finite temperatures. A first attempt to understand the effect of a
nonmagnetic defect on the properties of the $S=1/2$ 2D Heisenberg
antiferromagnet in the honeycomb lattice was made by
de Ch\^atel \emph{et al}.~\cite{dCC+04} In their mean-field approach, a single
impurity was introduced in clusters up to 12 spins. It is clear, however, that
their results can only be applied to systems with
dilutions up to $1-p=1/13$. Moreover, the random nature of defects
cannot be accounted for using their method.

In this paper, we follow the general idea of the work of Mucciolo \emph{et
al}.,~\cite{MNC04} by using the linear spin wave approximation  
in real space to compute different physical quantities. In addition 
we use finite-size scaling to determine the magnetic moment 
of the samples. We address the problem 
of determining the density of states (DOS) of our system  
using a different and more reliable method, 
which gives the behavior of the DOS in the thermodynamic
limit. The paper is organized as follows: in Sect.~\ref{sec:hamilt}
we present the Hamiltonian and the formalism we use; in Sect.~\ref{sec:numeric}
we give the numerical details of our method; we present the results
on the staggered magnetization and on density of states as well as
on the cluster characterization in Sect.~\ref{sec:results}; finally,
in Sect.~\ref{sec:conclusions} we summarize our work and present some
concluding remarks.

%%%%%%%%%%%%%%%%%%%%%%%%%%%%%%%%%%%%%%%%%%%%%%%%%%%%%%%%%%%%%%%%%%%%%%%%%%%%%%%
%%%%%%%%%%%%%%%%%%%%%%%%%%%%%%%%%%%%%%%%%%%%%%%%%%%%%%%%%%%%%%%%%%%%%%%%%%%%%%%
%%%%%%%%%%%%%%%%%%%%%%%%%%%%%%%%%%%%%%%%%%%%%%%%%%%%%%%%%%%%%%%%%%%%%%%%%%%%%%%
\section{Model Hamiltonian and formalism}
\label{sec:hamilt}

The Heisenberg Hamiltonian describing quantum spins in a site-diluted
honeycomb lattice is written as
\begin{equation}
H=J\sum_{i\in A,\bm\delta}\eta_{i}\eta_{i+\bm\delta}
\mathbf{S}_{i}^{a}\cdot\mathbf{S}_{i+\bm\delta}^{b}\,,
\label{eq:hamilt1}
\end{equation}
where $\mathbf{S}_{i}^a$ ($\mathbf{S}_{i}^b$) is the spin
operator on a site $i$ of sublattice $A$ ($B$). The notation $i+\bm\delta$
represents a nearest neighbor site of site $i$, connected to $i$
by the vector $\bm\delta$. There are three different $\bm\delta$
vectors given by
\begin{equation}
\bm\delta_{1}=\frac{c}{2}(1,\sqrt{3})\,,\hspace{0.25cm}
\bm\delta_{2}=\frac{c}{2}(1,-\sqrt{3})\,,\hspace{0.25cm}
\bm\delta_{3}=-c(1,0)\,,
\end{equation}
where $c$ is the hexagon side length. The $\eta_{i}$ variables
can have the values 0 or 1 depending on whether the site $i$ exists
or not.

The usual spin wave approximation starts by assuming that LRO exists and,
in the case of antiferromagnetism, that the ground state is not substantially
different from the N\'{e}el state. The mathematical meaning of this similarity
is that the following inequalities should hold:
\begin{alignat}{2}
S-\bigl\langle S_{i}^{a,z}\bigr\rangle
 \ll S & \quad & \text{for $i\in A$,}\label{eq:NsGA}\\
S+\bigl\langle S_{i}^{b,z}\bigr\rangle
 \ll S & \quad & \text{for $i\in B$.} \label{eq:NsGB}
\end{alignat}
With these in mind we express the spin operators in terms of bosonic
creation and annihilation operators as introduced by Holstein and
Primakoff.\cite{HP40} Holstein-Primakoff transformation is defined
for sublattice $A$ as 
\begin{equation} \label{eq:sopa}
\begin{split}
S_{i}^{a,z} & = S-a_{i}^{\dagger}a_{i}\,, \\
S_{i}^{a,+} & = \sqrt{2S}\sqrt{1-\frac{a_{i}^{\dagger}a_{i}}{2S}}\, a_{i}\,, \\
S_{i}^{a,-} & = \sqrt{2S}\, a_{i}^{\dagger}
 \sqrt{1-\frac{a_{i}^{\dagger}a_ {i}}{2S}}\,.
\end{split}
\end{equation}
In sublattice $B$ the spin have $S_{z}=-S$ projection in the N\'{e}el
state. Since the bosons should describe excitations above the ground
state, and this has to be such that inequalities \eqref{eq:NsGA}
and \eqref{eq:NsGB} are verified, the $S^{b,z}$ operator needs to
be redefined as $S_{i}^{b,z}=-S+b_{i}^{\dagger}b_{i}$. Accordingly,
the $S_{i}^{b,+}$ operator must create bosons, and all the operators
in sublattice $B$ are defined as 
\begin{equation} \label{eq:sopb}
\begin{split}
S_{i}^{b,z} & = -S+b_{i}^{\dagger}b_{i}\,, \\
S_{i}^{b,-} & = \sqrt{2S}\sqrt{1-\frac{b_{i}^{\dagger}b_{i}}{2S}}\, b_{i}\,, \\
S_{i}^{b,+} & = \sqrt{2S}\, b_{i}^{\dagger}
 \sqrt{1-\frac{b_{i}^{\dagger}b_{i}}{2S}}\,.
\end{split}
\end{equation}
It is worth mentioning that inequalities \eqref{eq:NsGA} and \eqref{eq:NsGB}
can also be expressed in terms of the new bosonic operators $a$ and
$b$ as 
\begin{alignat}{2}
\bigl\langle a_{i}^{\dagger}a_{i}\bigr\rangle
 \ll S & \quad & \text{for $i\in A$,} \label{eq:ineqAb} \\
\bigl\langle b_{i}^{\dagger}b_{i}\bigr\rangle
 \ll S & \quad & \text{for $i\in B$,} \label{eq:ineqBb}
\end{alignat}
from which the linear spin wave approximation follows straightforwardly
by expanding the square roots in Eqs.~\eqref{eq:sopa} and \eqref{eq:sopb}
in powers of $1/S$ and keeping only the zeroth order terms:
\begin{equation} \label{eq:Sapprox}
\begin{aligned}
S_{i}^{a,+} & \simeq \sqrt{2S}\, a_{i}\,, \\
S_{i}^{a,-} & \simeq \sqrt{2S}\, a_{i}^{\dagger}\,,
\end{aligned}
\quad
\begin{aligned}
S_{i}^{b,-} & \simeq \sqrt{2S}\, b_{i}\,, \\
S_{i}^{b,+} & \simeq \sqrt{2S}\, b_{i}^{\dagger}\,, 
\end{aligned}
\end{equation}
Inserting the resultant approximation \eqref{eq:Sapprox} into
Eq.~\eqref{eq:hamilt1} produces the linear spin wave Hamiltonian,
which reads 
\begin{multline} \label{eq:haf}
H = -Jh_{a}S(S+1)\sum_{i\in A,\bm\delta}\eta_{i}\eta_{i+\bm\delta} \\
+ JS\sum_{i\in A,\bm\delta}\eta_{i}\eta_{i+\bm\delta}\Bigl[h_{a}
\bigl(a_{i}a_{i}^{\dagger}+b_{i+\bm\delta}^{\dagger}b_{i+\bm\delta}\bigr) \\
+  a_{i}b_{i+\bm\delta}+b_{i+\bm\delta}^{\dagger}a_{i}^{\dagger}\Bigr]\,.
\end{multline}
Note that we have introduced a magnetic anisotropy $h_{a}$ in the
$S_{i}^{a,z}S_{i+\bm{\delta}}^{b,z}$ term.

The linear spin wave Hamiltonian \eqref{eq:haf} can be seen as having
a classical part of the form
\begin{equation}
H_{\text{cl}}=-Jh_{a}S(S+1)\sum_{i\in A,\bm\delta}\eta_{i}\eta_{i+\bm\delta}\,,
\label{eq:Hcl}
\end{equation}
and a quantum fluctuating part, which can be written as 
\begin{equation}
H_{\text{sw}}=(\{ a\},\{ b^{\dagger}\})\mathbf{D}
 (\{ a\},\{ b^{\dagger}\})^{\dagger}\,,
\label{eq:Hsw}
\end{equation}
where $(\{ a\},\{ b^{\dagger}\})^{\dagger}$ is a column
vector containing all the boson operators and
\begin{equation} \label{eq:matrixD}
\mathbf{D}=\begin{pmatrix}
\mathbf{K}^{a} & \mathbf{\Delta}\\
\mathbf{\Delta}^{T} & \mathbf{K}^{b}
\end{pmatrix}
\end{equation}
is the so-called grand dynamical matrix. For a diluted lattice, 
the number of sites in sublattice $A$ need not be the same as that
in sublattice $B$; therefore the dimensions of the blocks in
$\mathbf{D}$ are $N_{a}\times N_{a}$ for $\mathbf{K}^{a}$,
$N_{b}\times N_{b}$ for $\mathbf{K}^{b}$, $N_{a}\times N_{b}$ for
$\mathbf{\Delta}$ and $N_{b}\times N_{a}$ for $\mathbf{\Delta}^{T}$.
The corresponding matrix elements are
\begin{flalign}
\quad K_{ij}^{a}&=
 Jh_{a}S\delta_{ij}\eta_{i}\sum_{\bm\delta}\eta_{i+\bm\delta}\,,
& \text{for $i\in A$,} \label{eq:Ka}\\
\quad K_{ij}^{b}&=
 Jh_{a}S\delta_{ij}\eta_{i}\sum_{\bm\delta}\eta_{i+\bm\delta}\,,
& \text{for $i\in B$,} \label{eq:Kb}\\
\quad \Delta_{ij}&=JS\eta_{i}\eta_{j}\,,
& \text{for $i\in A$, $j\in i+\bm\delta$,} \\
\quad \Delta_{ij}^{T}&=JS\eta_{i}\eta_{j}\,,
& \text{for $i\in B$, $j\in i+\bm\delta$.}  \label{eq:DeltaT}
\end{flalign}

The diagonalization of the bosonic Hamiltonian amounts to find a transformation
$\mathbf{T}$ such that
\begin{equation}
(\mathbf{T}^{\dagger})^{-1}\mathbf{D}\mathbf{T}^{-1}=
 \diag(\omega_{1},\ldots,\omega_{N_{a}},\omega_{N_{a}+1},\ldots,
 \omega_{N_{a}+N_{b}})\,,
\label{eq:diag}
\end{equation}
where $\diag(\omega_{1},\ldots, \omega_{N_a+N_b})$ stands for a diagonal matrix
with elements $\omega_{1},\ldots, \omega_{N_a+N_b}$ in its diagonal,
$N_{a}+N_{b}$ in number.
In this case all the eigenvalues $\omega_{1},\ldots,\omega_{N_{a}},
\omega_{N_{a}+1},\ldots,\omega_{N_{a}+N_{b}}$ are positive.
The quasi-particles associated with those eigenvalues are obtained from
\begin{equation}
(\{\alpha\},\{\beta^{\dagger}\})^{\dagger}=
 \mathbf{T}(\{ a\},\{ b^{\dagger}\})^{\dagger}\,.
\label{eq:rot}
\end{equation}

In the undiluted case, it is well known that Eq.~\eqref{eq:Hsw} can
be diagonalized through a Bogoliubov-Valatin transformation in the
reciprocal space. For a subsequent analysis it is convenient to reproduce
here the results of the calculation for the undiluted honeycomb
lattice.\cite{PAB04} We first introduce the operators $a_{\mathbf{k}}$
and $b_{\mathbf{k}}$ defined as the inverse Fourier transforms of
$a_{i}$ and $b_{i}$,
\begin{equation}
a_{i}=\frac{1}{\sqrt{N_{a}}}
 \sum_{\mathbf{k}}e^{-i\mathbf{k}\cdot\mathbf{r}_{i}}a_{\mathbf{k}}\,,
\hspace{0.21cm}
b_{i}=\frac{1}{\sqrt{N_{b}}}
\sum_{\mathbf{k}}e^{-i\mathbf{k}\cdot\mathbf{r}_{i}}b_{\mathbf{k}}\,,
\label{eq:akbk}
\end{equation}
where the $\mathbf{k}$ summation ranges over the first Brillouin
zone of either sublattice $A$ or $B$. (Do not confuse the site index $i$
and the complex imaginary unit also present in the Fourier
transform). The vector $\mathbf{r}_{i}$ is the position vector of
site $i$, and $N_{a}=N_{b}$ in the absence of dilution. Substituting
Eq.~\eqref{eq:akbk} into Hamiltonian~\eqref{eq:Hsw} gives us
$H_{\text{sw}}=\sum_{\mathbf{k}}H_{\mathbf{k}}$, with 
\begin{multline} \label{eq:Hk}
H_{\mathbf{k}}=JSz\Bigl[h_{a}\bigl(a_{\mathbf{k}}a_{\mathbf{k}}^{\dagger}
 + b_{-\mathbf{k}}^{\dagger}b_{-\mathbf{k}}\bigr)\\
+\phi_{\mathbf{k}}a_{\mathbf{k}}b_{-\mathbf{k}}
 +\phi_{\mathbf{k}}^{*}b_{-\mathbf{k}}^{\dagger}a_{\mathbf{k}}^{\dagger}\Bigr],
\end{multline}
 where $\phi_{\mathbf{k}}$ is defined as
\begin{equation}
\phi_{\mathbf{k}}=\frac{1}{z}\sum_{\bm\delta}e^{-i\mathbf{k}\cdot\bm\delta}\,.
\label{eq:phik}
\end{equation}
The diagonalized form of Hamiltonian \eqref{eq:Hk}, given by 
\begin{equation} \label{eq:Hkdiag}
H_{\mathbf{k}}=\omega_{\mathbf{k}}
 \bigl(1+\alpha_{\mathbf{k}}^{\dagger}\alpha_{\mathbf{k}}
 +\beta_{\mathbf{k}}^{\dagger}\beta_{\mathbf{k}}\bigr)\,,
\end{equation}
with
\begin{equation}
\omega_{\mathbf{k}}=JSz\sqrt{h_{a}^{2}-\left|\phi_{\mathbf{k}}\right|^{2}},
\label{eq:Ek}
\end{equation}
can be easily obtained from the following Bogoliubov-Valatin transformation,
\begin{equation} \label{eq:BVT}
\begin{split}
\alpha_{\mathbf{k}} & = 
 u_{\mathbf{k}}a_{\mathbf{k}}+v_{\mathbf{k}}b_{-\mathbf{k}}^{\dagger}\,, \\
\beta_{\mathbf{k}} & = 
 v_{\mathbf{k}}a_{\mathbf{k}}^{\dagger}+u_{\mathbf{k}}b_{-\mathbf{k}}\,,
\end{split}
\end{equation}
with coefficients $u_{\mathbf{k}}$ and $v_{\mathbf{k}}$
given as functions of the parameters $h_{a}$ and $\phi_{\mathbf{k}}$.

In the diluted case, translational symmetry is lost, and the solution
in the reciprocal space is as difficult as the one in real space. Let us start
by constructing an operator transformation of the Bogoliubov-Valatin
type in real space which can be used in the presence of dilution:
\begin{align} 
\alpha_{n}&=\sum_{i=1}^{N_{a}}u_{ni}a_{i}+
\sum_{i=1}^{N_{b}}v_{ni}b_{i}^{\dagger}\,, \label{eq:alpha}\\
\beta_{n}&=\sum_{i=1}^{N_{a}}w_{ni}a_{i}^{\dagger}+
 \sum_{i=1}^{N_{b}}x_{ni}b_{i}\,.\label{eq:beta}
\end{align}
This definition of $\alpha^{\dagger}$ and $\beta^{\dagger}$ excitations
gives us $N_{a}$ $\alpha$-type quasi-particles and $N_{b}$
$\beta$-type quasi-particles. Equations \eqref{eq:alpha}
and \eqref{eq:beta} define the transformation matrix \eqref{eq:rot},
\begin{equation} \label{eq:T}
\mathbf{T}=\begin{pmatrix}
\mathbf{U}^{*} & \mathbf{V}^{*}\\
\mathbf{W} & \mathbf{X}\end{pmatrix},
\end{equation}
 where the $N_{a}\times N_{a}$ ($N_{b}\times N_{a}$) and $N_{a}\times N_{b}$
($N_{b}\times N_{b}$) matrices $\mathbf{U}^{*}$ ($\mathbf{W}$)
and $\mathbf{V}^{*}$ ($\mathbf{X}$) contain the coefficients $u_{ni}^{*}$
($w_{ni}$) and $v_{ni}^{*}$ ($x_{ni}$), respectively. Since the
quasi-particles must have a bosonic character, the quasi-particle operators
must satisfy the commutation relations
\begin{align}
\bigl[\alpha_{n},\alpha_{m}^{\dagger}\bigr]=
 \bigl[\beta_{n},\beta_{m}^{\dagger}\bigr] & = \delta_{nm}\,,\\
\bigl[\alpha_{n},\beta_{m}\bigr]=
 \bigl[\beta_{n}^{\dagger},\alpha_{m}^{\dagger}\bigr] & = 0\,,
\end{align}
which lead to the following constraints on the transformation coefficients:
\begin{align}
\sum_{i=1}^{N_{a}}u_{ni}u_{mi}^{\ast}-\sum_{i=1}^{N_{b}}v_{ni}v_{mi}^{\ast}
 & = \delta_{nm}\,,\label{eq:ORalpha}\\
\sum_{i=1}^{N_{a}}w_{ni}w_{mi}^{\ast}-\sum_{i=1}^{N_{b}}x_{ni}x_{mi}^{\ast}
 & = -\delta_{nm}\,,\label{eq:ORbeta}\\
\sum_{i=1}^{N_{a}}u_{ni}w_{mi}-\sum_{i=1}^{N_{b}}v_{ni}x_{mi}
 & = 0\,,\label{eq:ORaphbet}\\
\sum_{i=1}^{N_{a}}w_{ni}^{\ast}u_{mi}^{\ast}-
 \sum_{i=1}^{N_{b}}x_{ni}^{\ast}v_{mi}^{\ast} & = 0\,.
\label{eq:ORalphbetcc}
\end{align}

Equations~(\ref{eq:ORalpha}--\ref{eq:ORalphbetcc}) can be written
in matrix notation as
\begin{align}
\mathbf{UU}^{\dagger}-\mathbf{VV}^{\dagger}=\mathbf{U}^{*}\mathbf{U}^{T}-
 \mathbf{V}^{*}\mathbf{V}^{T} & = \mathbf{I}_{N_{a}} \label{eq:UUVV}\,,\\
\mathbf{W}\mathbf{W}^{\dagger}-\mathbf{X}\mathbf{X}^{\dagger}=
 \mathbf{W}^{*}\mathbf{W}^{T}-\mathbf{X}^{*}\mathbf{X}^{T} 
 & =  -\mathbf{I}_{N_{b}}\,,\\
\mathbf{UW}^{T}-\mathbf{VX}^{T}=\mathbf{WU}^{T}-\mathbf{XV}^{T}
 & = \mathbf{0}\,,\\
\mathbf{W}^{*}\mathbf{U}^{\dagger}-\mathbf{X}^{*}\mathbf{V}^{\dagger}=
 \mathbf{U}^{*}\mathbf{W}^{\dagger}-\mathbf{V}^{*}\mathbf{X}^{\dagger}
 & = \mathbf{0}\,,
\label{eq:WUXV}
\end{align}
where $\mathbf{I}_{N_{a}}$ ($\mathbf{I}_{N_{b}}$) is the $N_{a}\times N_{a}$
($N_{b}\times N_{b}$) unit matrix. These relations can be put into
a more compact form by defining the matrix 
\begin{equation}
\mathbf{1}_{p}=\begin{pmatrix}
\mathbf{I}_{N_{a}} & \mathbf{0}\\
\mathbf{0} & -\mathbf{I}_{N_{b}}\end{pmatrix},
\label{eq:1p}
\end{equation}
in terms of which Eqs.~(\ref{eq:UUVV}--\ref{eq:WUXV}) can be rewritten as
\begin{equation}
\mathbf{T}\mathbf{1}_{p}\mathbf{T}^{\dagger}=\mathbf{1}_{p}\,.
\label{eq:T1pT+}
\end{equation}
Since $\mathbf{1}_{p}\mathbf{1}_{p}=\mathbf{I}_{N_{a}+N_{b}}$, it
can be shown, after simple algebraic transformations, that
\begin{equation}
\mathbf{T}^{\dagger}\mathbf{1}_{p}\mathbf{T}=\mathbf{1}_{p}\,.
\label{eq:T+1pT}\end{equation}
As a result we have four additional (though not independent) sets
of orthogonality equations, 
\begin{align}
\sum_{n=1}^{N_{a}}u_{ni}u_{nj}^{\ast}-\sum_{n=1}^{N_{b}}w_{ni}^{\ast}w_{nj}
 & = \delta_{ij}\,,\\
\sum_{n=1}^{N_{a}}v_{ni}v_{nj}^{\ast}-\sum_{n=1}^{N_{b}}x_{ni}^{\ast}x_{nj}
 & = -\delta_{ij}\,,\\
\sum_{n=1}^{N_{a}}v_{ni}u_{nj}^{\ast}-\sum_{n=1}^{N_{b}}x_{ni}^{\ast}w_{nj}
 & = 0\,,\\
\sum_{n=1}^{N_{a}}u_{ni}v_{nj}^{\ast}-\sum_{n=1}^{N_{b}}w_{ni}^{\ast}x_{nj}
 & = 0\,.
\end{align}
Since transformations \eqref{eq:alpha} and \eqref{eq:beta} diagonalize
the spin wave Hamiltonian \eqref{eq:Hsw}, the quasi-particles will
obey the following commutation relations:
\begin{align}
\left[\alpha_{n},H_{\text{sw}}\right]
 & = \omega_{n}^{(\alpha)}\alpha_{n}\,,\label{eq:alphaH}\\
\left[\beta_{n},H_{\text{sw}}\right] & = \omega_{n}^{(\beta)}\beta_{n}\,,\\
\left[\alpha_{n}^{\dagger},H_{\text{sw}}\right]
 & = -\omega_{n}^{(\alpha)}\alpha_{n}^{\dagger}\,,\\
\left[\beta_{n}^{\dagger},H_{\text{sw}}\right]
 & = -\omega_{n}^{(\beta)}\beta_{n}^{\dagger}\,.\label{eq:betapH}
\end{align}
[Notice that we have relabeled the positive eigenvalues introduced
in \eqref{eq:diag} to $\omega_{1}^{(\alpha)},\ldots,\omega_{N_{a}}^{(\alpha)},
\omega_{1}^{(\beta)},\ldots,\omega_{N_{b}}^{(\beta)}$.]
From Eqs.~\eqref{eq:alphaH} and \eqref{eq:alpha} we can define an eigenvalue
matrix equation in the usual form, namely, 
\begin{equation}
\begin{pmatrix}
\mathbf{K}^{a} & -\mathbf{\Delta}\\
\mathbf{\Delta}^{T} & -\mathbf{K}^{b}\end{pmatrix}
\begin{pmatrix}
\bar{u}_{n}\\
\bar{v}_{n}\end{pmatrix}
=\omega_{n}^{(\alpha)}
\begin{pmatrix}
\bar{u}_{n}\\
\bar{v}_{n}
\end{pmatrix},
\label{eq:eiguv}
\end{equation}
where the column vectors $\bar{u}_{n}$ and $\bar{v}_{n}$ contain
the coefficients $u_{ni}$ and $v_{ni}$, respectively. From
Eq.~\eqref{eq:betapH} and the complex conjugate of Eq.~\eqref{eq:beta},
a similar eigenvalue matrix equation can be defined, 
\begin{equation} \label{eq:eigwx}
\begin{pmatrix}
\mathbf{K}^{a} & -\mathbf{\Delta}\\
\mathbf{\Delta}^{T} & -\mathbf{K}^{b}
\end{pmatrix}
\begin{pmatrix}
\bar{w}_{n}^{\ast}\\
\bar{x}_{n}^{\ast}
\end{pmatrix}
=-\omega_{n}^{(\beta)}
\begin{pmatrix}
\bar{w}_{n}^{\ast}\\
\bar{x}_{n}^{\ast}
\end{pmatrix}.
\end{equation}
Defining the matrices
\begin{equation}
\mathbf{\Omega}_{\alpha}=
 \diag(\omega_{1}^{(\alpha)},\ldots,\omega_{N_{a}}^{(\alpha)})\,,
\end{equation}
and
\begin{equation}
\mathbf{\Omega}_{\beta}=
 \diag(\omega_{1}^{(\beta)},\ldots,\omega_{N_{b}}^{(\beta)})\,,
\end{equation}
and recalling definition \eqref{eq:T} for $\mathbf{T}$ and definition
\eqref{eq:1p} for $\mathbf{1}_{p}$, Eqs.~\eqref{eq:eiguv} and \eqref{eq:eigwx}
can be combined into a single equation, 
\begin{equation}\label{eq:eigprob}
\mathbf{D}\mathbf{1}_{p}\mathbf{T}^{\dagger}=\mathbf{T}^{\dagger}
\begin{pmatrix}
\mathbf{\Omega}_{\alpha} & \mathbf{0}\\
\mathbf{0} & -\mathbf{\Omega}_{\beta}
\end{pmatrix}.
\end{equation}
This can be made more compact still by defining the matrix 
$\mathbf{\Omega}=\diag(\omega_{1}^{(\alpha)},\ldots,
\omega_{N_{a}}^{(\alpha)},\omega_{1}^{(\beta)},\ldots,
\omega_{N_{b}}^{(\beta)})$, such that
\begin{equation}
\mathbf{D}\mathbf{1}_{p}\mathbf{T}^{\dagger}=
 \mathbf{T}^{\dagger}\mathbf{\Omega}\mathbf{1}_{p}\,.
\label{DST-TSO}
\end{equation}
From Eq.~\eqref{DST-TSO} and the relations \eqref{eq:T1pT+} and
\eqref{eq:T+1pT}, it is not difficult to show that
\begin{equation}
\mathbf{1}_{p}\mathbf{T}\mathbf{1}_{p}\mathbf{D}
 \mathbf{1}_{p}\mathbf{T}^{\dagger}\mathbf{1}_{p}=\mathbf{\Omega}\,.
\end{equation}
Thus, solving the eigenvalue problem defined by Eq.~\eqref{eq:eigprob}
under the constraint \eqref{eq:T1pT+} is equivalent to finding a transformation
$\mathbf{T}$ which satisfies Eq.~\eqref{eq:diag}, and where, obviously,
\begin{equation}
\mathbf{1}_{p}\mathbf{T}^{\dagger}\mathbf{1}_{p}=\mathbf{T}^{-1}\,.
\end{equation}
According to Eq.~\eqref{eq:rot}, the diagonalized form of the spin wave
Hamiltonian is obtained as 
\begin{equation}
\begin{split}
H_{\text{sw}} & = (\{ a\},\{ b^{\dagger}\})\mathbf{D}
 (\{ a\},\{ b^{\dagger}\})^{\dagger}\nonumber \\
 & = (\{\alpha\},\{\beta^{\dagger}\})\mathbf{\Omega}
 (\{\alpha\},\{\beta^{\dagger}\})^{\dagger}\nonumber \\
 & = \sum_{n=1}^{N_{a}}\omega_{n}^{(\alpha)}\alpha_{n}\alpha_{n}^{\dagger}
 +\sum_{n=1}^{N_{b}}\omega_{n}^{(\beta)}\beta_{n}^{\dagger}\beta_{n}\,.
\end{split}
\end{equation}

The conclusion of the above discussion is that the operator transformation
given by Eqs.~\eqref{eq:alpha} and \eqref{eq:beta} diagonalizes
the spin wave Hamiltonian \eqref{eq:haf} and that $\alpha^{\dagger}$
and $\beta^{\dagger}$ are the quasi-particles (with bosonic character)
associated with the low energy excitations of the antiferromagnetic
Heisenberg Hamiltonian for quantum spins in a site diluted honeycomb
lattice (needless to say the above description applies to other lattices
as well).

Using Eq.~\eqref{eq:rot} it is possible to write any average of the
initial bosonic operators in terms of the quasi-particle operators
$\alpha$ and $\beta$. The simplest example is the staggered magnetization
$M_{z}^{\text{stagg}}$ at $T=0$ given by
\begin{equation} \label{eq:magz}
\begin{split}
M_{z}^{\text{stagg}} & =  \Biggl\langle \sum_{i\in A}S_{i}^{a,z}
 -\sum_{i\in B}S_{i}^{b,z}\Biggr\rangle \\
& = \bigl(N_{a}+N_{b}\bigr)\bigl(S-\delta m_{z}\bigr)\,,
\end{split}
\end{equation}
where
\begin{equation} \label{eq:deltamz}
\delta m_{z}=\sum_{n=1}^{N_{a}}\delta m_{z}^{(n,\alpha)}
 +\sum_{n=1}^{N_{b}}\delta m_{z}^{(n,\beta)}\,,
\end{equation}
with
\begin{align} 
\delta m_{z}^{(n,\alpha)}
 & = \frac{1}{N_{a}+N_{b}}\sum_{i\in B}|v_{ni}|^{2}\,, \label{eq:deltamzalph}
\intertext{and}
\delta m_{z}^{(n,\beta)}
 & = \frac{1}{N_{a}+N_{b}}\sum_{i\in A}|w_{ni}|^{2}.\label{eq:deltamzbet}
\end{align}

%%%%%%%%%%%%%%%%%%%%%%%%%%%%%%%%%%%%%%%%%%%%%%%%%%%%%%%%%%%%%%%%%%%%%%%%%%%%%%%
%%%%%%%%%%%%%%%%%%%%%%%%%%%%%%%%%%%%%%%%%%%%%%%%%%%%%%%%%%%%%%%%%%%%%%%%%%%%%%%
%%%%%%%%%%%%%%%%%%%%%%%%%%%%%%%%%%%%%%%%%%%%%%%%%%%%%%%%%%%%%%%%%%%%%%%%%%%%%%%
\section{Numerical details}
\label{sec:numeric}

The formalism developed in the above section is based on the existence
of the matrix $\mathbf{T}$ and, naturally, on the possibility of
finding it by some numerical procedure. In this work we have used two
independent methods to compute the transformation matrix and the associated
eigenenergies. Both methods agree with each other within the numerical
accuracy of the calculation. One of them is based on a Cholesky decomposition
and gives the $\mathbf{T}^{-1}$ matrix directly, whereas the other
solves the eigenvalue problem defined by Eq.~\eqref{eq:eigprob},
computing the matrix $\mathbf{T}^{\dagger}$ and from this  the
matrix $\mathbf{T}$.

%%%%%%%%%%%%%%%%%%%%%%%%%%%%%%%%%%%%%%%%%%%%%%%%%%%%%%%%%%%%%%%%%%%%%%%%%%%%%%%
%%%%%%%%%%%%%%%%%%%%%%%%%%%%%%%%%%%%%%%%%%%%%%%%%%%%%%%%%%%%%%%%%%%%%%%%%%%%%%%
\subsection{Cholesky Decomposition method}
\label{Cholesky}

As shown by Colpa,\cite{colpa} so long as the grand dynamical matrix
is positive definite, a simple algorithm exists for determining $\mathbf{T}$.
A hermitian (or symmetric) matrix is positive definite if all its
eigenvalues are positive. By definition the quasi-particles $\alpha^{\dagger}$
and $\beta^{\dagger}$ have positive or zero excitation energy. As
will be shown in Subsect.~\ref{sub:Zm}, the zero energy excitations
are associated with spin rotations, which cost zero energy due to
the spin rotational symmetry of the isotropic Heisenberg model. So,
provided that $h_{a}\geq1^{+}$, all eigenvalues are positive and the
grand dynamical matrix is positive definite. The algorithm is implemented
in three major steps:
\begin{enumerate}
\item for $\mathbf{D}$ positive definite a Cholesky decomposition can be
performed \cite{recipes} and we have 
$\mathbf{D}=\mathbf{Q}\mathbf{Q}^{\dagger}$,
where $\mathbf{Q}$ is an upper triangular matrix. The existence of
a Cholesky decomposition guarantees that the problem is positive definite; 
\item it can be proved that there exists a unitary transformation $\mathbf{Y}$
such that $\mathbf{Y}^{\dagger}(\mathbf{Q}\mathbf{1}_{p}^{(ab)}
\mathbf{Q}^{\dagger})\mathbf{Y}=\mathbf{1}_{p}^{(ab)}\diag(\omega_{1},
\ldots,\omega_{N_{a}},\omega_{N_{a}+1},\ldots\omega_{N_{a}+N_{b}})$;
\item finally, it can be proved that $\mathbf{T}^{-1}=\mathbf{Q}^{-1}
\mathbf{Y}\diag(\sqrt{\omega_{1}},\ldots,\sqrt{\omega_{N_{a}+N_{b}}})$.
\end{enumerate}

%%%%%%%%%%%%%%%%%%%%%%%%%%%%%%%%%%%%%%%%%%%%%%%%%%%%%%%%%%%%%%%%%%%%%%%%%%%%%%%
%%%%%%%%%%%%%%%%%%%%%%%%%%%%%%%%%%%%%%%%%%%%%%%%%%%%%%%%%%%%%%%%%%%%%%%%%%%%%%%
\subsection{Bogoliubov-Valatin Transformation method}
\label{sub:BV}

The nonhermitian eigenproblem defined by Eq.~\eqref{eq:eigprob}
can be solved with standard numerical algorithms. Here we have used
subroutines of the LAPACK library. It should be noted that the resultant
eigenvectors do not provide the required $\mathbf{T}^{\dagger}$ matrix
directly. After diagonalization the eigenvectors have to be normalized
such that they satisfy Eqs.~\eqref{eq:ORalpha} and \eqref{eq:ORbeta}
for $n=m$. Degenerate eigenvectors%
\footnote{Although degeneracy is removed by disorder, we have to handle it if
we want our algorithm to be valid in the undiluted case.%
} have to be carefully analyzed because the LAPACK subroutines we have
used do not guarantee that they satisfy Eq.~\eqref{eq:T1pT+}
(though Eqs.~\eqref{eq:ORalpha} to \eqref{eq:ORalphbetcc} are satisfied
by default for $n\neq m$). The algorithm is implemented as follows:
\begin{enumerate}
\item the matrix $\mathbf{D}\mathbf{1}_{p}^{(ab)}$ is reduced to an upper
Hessenberg form $\mathbf{H}$ by an orthogonal transformation $\mathbf{Y}$,
\emph{i.e.}, $\mathbf{H}=\mathbf{YD1}_{p}^{(ab)}\mathbf{Y}^{\dagger}$ (LAPACK
subroutines DGEHRD and DORGHR); 
\item the eigenvalues of the upper Hessenberg matrix (the same as those
of $\mathbf{D}\mathbf{1}_{p}^{(ab)}$) and the matrices $\mathbf{Q}$
and $\mathbf{Z}$ from the Schur decomposition $\mathbf{H}=
\mathbf{ZQZ}^{\dagger}$, where $\mathbf{Q}$ is an upper quasi-triangular
matrix (the Schur form), and $\mathbf{Z}$ is the orthogonal matrix of Schur
vectors, are computed (LAPACK subroutine DHSEQR);
\item the right eigenvectors of the upper quasi-triangular matrix $\mathbf{Q}$
are computed and multiplied by $\mathbf{Y}^{\dagger}\mathbf{Z}$,%
\footnote{Subroutine DHSEQR already gives the product
$\mathbf{Y}^{\dagger}\mathbf{Z}$ instead of $\mathbf{Z}$.%
} giving the right eigenvectors of $\mathbf{D}\mathbf{1}_{p}^{(ab)}$,
whose matrix form we name $\mathbf{\widetilde{T}}^{\dagger}$ (LAPACK
subroutines DTREVC); 
\item degenerate column eigenvectors of $\mathbf{\widetilde{T}}^{\dagger}$
are arranged in linear combinations such that they satisfy 
Eq.~\eqref{eq:T1pT+};
\item nondegenerate column eigenvectors of $\mathbf{\widetilde{T}}^{\dagger}$
are normalized so as to satisfy the orthogonality relations of Eqs.
\eqref{eq:ORalpha} and \eqref{eq:ORbeta}, giving matrix
$\mathbf{T}^{\dagger}$;
\item the positive eigenvalues and respective eigenvectors are identified
with $\alpha$ modes, and the negative ones with the $\beta$ modes; 
\item eigenvalues and eigenvectors are sorted such that matrix
$\mathbf{T}^{\dagger}$ has the form defined in Eq.~\eqref{eq:T}. 
\end{enumerate}

%%%%%%%%%%%%%%%%%%%%%%%%%%%%%%%%%%%%%%%%%%%%%%%%%%%%%%%%%%%%%%%%%%%%%%%%%%%%%%%
%%%%%%%%%%%%%%%%%%%%%%%%%%%%%%%%%%%%%%%%%%%%%%%%%%%%%%%%%%%%%%%%%%%%%%%%%%%%%%%
\subsection{Zero modes}
\label{sub:Zm}

It is well known that the clean and isotropic limit of Hamiltonian
\eqref{eq:haf} has two zero-energy excitations (Goldstone bosons),
whose momenta can be determined from Eq.~\eqref{eq:Hkdiag}. These
gapless modes are a consequence of the fact that the ground state
spontaneously breaks the rotational symmetry of the Hamiltonian in
spin space. It can be shown\cite{ConcSolid63} that these zero-energy
modes have divergent amplitudes. In two and three dimensions the quantum
corrections to the staggered magnetization (at zero temperature) are
finite, meaning that the divergence associated with the zero energy
modes is integrable. We note, however, that if the mean square amplitudes
of the differences between the two $x$- and $y$-components, given by
\begin{align}
\frac{1}{N_{a}+N_{b}}&\left(\sum_{i\in A}S_{i}^{a,x}
 -\sum_{i\in B}S_{i}^{b,x}\right)\label{eq:magx}
\intertext{and}
\frac{1}{N_{a}+N_{b}}&\left(\sum_{i\in A}S_{i}^{a,y}
 -\sum_{i\in B}S_{i}^{b,y}\right)\,,\label{eq:magy}
\end{align}
are computed, we immediately find  divergent behavior.\cite{AND52}
The same divergent behavior is also found if we try to get the staggered
magnetization \eqref{eq:magz} from a finite-size scaling procedure,
doing summations in $\mathbf{k}\text{-space}$ for finite-size systems,
including all momenta of the Brillouin zone.

As shown by Anderson,\cite{AND52,ConcSolid63} this does not mean that
the spin wave approximation is breaking down and that the system
has no LRO. What it means is that these divergences are related to
the zero point motion of the Goldstone modes, and their presence is
required to exist since in a finite-size system one cannot have solutions
that break the spin rotational symmetry of the problem.\cite{Hulthen1938}
The presence of a \emph{broken symmetry ground state} is made possible
if we analyze the $H_{\mathbf{0}}$ term in Eq.~\eqref{eq:Hk}, from
which the Goldstone modes arise. This term cannot be diagonalized
through any Bogoliubov-Valatin transformation. Actually, it has a
continuum spectrum starting from the zero energy ground state (see
Appendix \ref{sec:App-diagHk0}). Using this continuum of states we
can form a wave packet centered around some prefixed orientation in
spin space, with the property of having both a finite staggered magnetization,
and a mean square roots of the quantities \eqref{eq:magx} and \eqref{eq:magy}
scaling with $1/N^{\frac{1}{2}-\alpha}$, with $\alpha>0$, as long
as we pay some extra energy. In addition, it can be shown (see Appendix
\ref{sec:App-Anderson}) that this extra energy scales as
$1/N^{\frac{1}{2}+\alpha}$, being negligible in the thermodynamic limit.
Thus it is a suitable approximation to form the above mentioned wave packet
from the solutions of $H_{\mathbf{0}}$, and to study the energy and the
zero point motion of all other normal modes within a time interval smaller
than that needed for the zero-energy wave packet to disrupt the coherence of
the unidirectional state.\cite{ConcSolid63} The understanding of
the role played by Goldstone bosons in finite-size systems is crucial
for computing well defined quantities in the thermodynamic limit from
calculations in finite-size lattices. As a practical example, the
staggered magnetization can be obtained from the finite-size calculations
if the Goldstone zero point motion contributions are subtracted, because
the $H_{\mathbf{0}}$ solutions were already used to form the starting
broken symmetry state. From this procedure we get exactly the same
value as from the convergent integral in the continuum limit.

The above discussion now needs to be carried on to the diluted case,
where the above aspects are more delicate than in the nondisordered
case. In the presence of dilution, it is easy to verify that there
is at least one zero mode in Eq.~\eqref{eq:eiguv} in the isotropic
case. This nontrivial solution with zero energy satisfies the equation
\begin{equation} \label{eq:zm}
\begin{pmatrix}
\mathbf{K}^{a} & -\mathbf{\Delta}\\
\mathbf{\Delta}^{T} & -\mathbf{K}^{b}
\end{pmatrix}
\begin{pmatrix}
\bar{c}\\
\bar{c}\end{pmatrix} =0,
\end{equation}
with all the amplitudes constant. To prove that this is indeed an
eigenstate we only need to remember definitions 
(\ref{eq:Ka}--\ref{eq:DeltaT}) of matrices $\mathbf{K}^{a}$, $\mathbf{K}^{b}$
and $\mathbf{\Delta}$, and check that the following equalities 
always hold:
\begin{align}
K_{ii}^{a} & = \!\sum_{j=N_{a}}^{N_{a}+N_{b}}\!\Delta_{ij}\,,\\
K_{ii}^{b} & = \sum_{j=1}^{N_{a}}\Delta_{ij}^{T}\,.
\end{align}

In terms of quasi-particle excitations, the eigenvector defined by
Eq.~\eqref{eq:zm} can be expressed as%
\footnote{The negligence of normalization doesn't change the conclusions we
will arrive. Actually, this modes will be identified with the Goldstone
modes of the diluted system, which, as in the clean limit, can have
divergent amplitude.}
\begin{align}
\alpha_{0}^{\dagger}&\propto\sum_{i=1}^{N_{a}}a_{i}^{\dagger}
 +\sum_{i=1}^{N_{b}}b_{i}\,,\label{eq:zm-alph-ab}
\intertext{in the case of an $\alpha$-type excitation, and}
\beta_{0}^{\dagger}&\propto\sum_{i=1}^{N_{a}}a_{i}
 +\sum_{i=1}^{N_{b}}b_{i}^{\dagger} \label{eq:zm-beta-ab}\,,
\end{align}
if it is a $\beta$-type excitation. Recalling the approximate
expressions in Eq.~\eqref{eq:Sapprox} for the operators $S_{i}^{(a,b)+}$
and $S_{i}^{(a,b)-}$ in terms of bosonic operators $a$ and $b$,
Eqs.~\eqref{eq:zm-alph-ab} and \eqref{eq:zm-beta-ab} can be rewritten as 
\begin{align}
\alpha_{0}^{\dagger} & \propto S_{\text{tot}}^{-}\,, \label{eq:zm-alph-Stot-}\\
\beta_{0}^{\dagger} & \propto S_{\text{tot}}^{+}\,. \label{eq:zm-beta-Stot+}
\end{align}
Thus, excitations $\alpha_{0}^{\dagger}$ and $\beta_{0}^{\dagger}$
are precisely the Goldstone bosons associated with the broken
continuous symmetry of spin rotation in the diluted system.

As will be shown in Subsect.~\ref{sub:Finite-size-scaling}, the thermodynamic
limit of the staggered magnetization for the diluted system will be
obtained from a finite-size scaling analysis. As we have started from
a broken symmetry ground state (the wave packet), which is a direct
consequence of Eqs.~\eqref{eq:NsGA} and \eqref{eq:NsGB}, we would
proceed as in the clean limit and neglect the contributions of $\alpha_{0}$
and $\beta_{0}$ modes. However, although in the undiluted case
the number of Goldstone modes is always two, when dilution is present
this number can either be one or two, in a finite size lattice. The
reason why this is so is that operators $S_{\text{tot}}^{-}$ and
$S_{\text{tot}}^{+}$ do not always represent independent excitations,
\emph{i.e.}, they do not always commute. Naturally $S_{\text{tot}}^{-}$
and $S_{\text{tot}}^{+}$ never commute strictly speaking because
\begin{equation}
[S_{\text{tot}}^{+},S_{\text{tot}}^{-}]=2S_{\text{tot}}^{z}\,.
\label{eq:commutS+S-}
\end{equation}
Nevertheless, in the clean limit we can easily convince ourselves
that the expectation value of $S_{\text{tot}}^{z}$ is always zero,
and, as $S_{\text{tot}}^{z}$ is a constant of the motion, commutator
\eqref{eq:commutS+S-} will always be zero. To get the value of the
commutator \eqref{eq:commutS+S-} in the presence of dilution we make
use of Eq.~\eqref{eq:Sapprox}, from which one finds
\begin{equation}
[S_{\text{tot}}^{+},S_{\text{tot}}^{-}]\propto N_{a}-N_{b}\,.
\label{eq:commutS+S-dilute}
\end{equation}
Now it is easily seen that one can have one or two Goldstone modes
in a finite-size diluted system: if the number of undiluted sites
in each sublattice is the same ($N_{a}=N_{b}$) there will be two
Goldstone modes; otherwise, if $N_{a}\neq N_{b}$, there will be only
one. Applying to this case the reasoning used for the undiluted
case, we should then neglect the contributions of the existent Goldstone
modes.

As the system size increases, the fluctuations relative to the zero
mean value of $N_{a}-N_{b}$ should scale as $1/\sqrt{N_{a}+N_{b}}$,
statistically speaking. Therefore the difference $N_{a}-N_{b}$ is
again zero in the thermodynamic limit and the system has two zero
energy excitations. This situtation cannot be achieved in finite
size lattices, unless we restrict ourselvess to cases where the disordered
realizations are constrained to obey the condition $N_{a}=N_{b}$, being
clear that the staggered magnetization in the thermodynamic limit
cannot depend on this restriction. We stress, however, that without
this restriction the conclusions drawn from finite-size lattices would
be different if we had accepted all sorts of disordered lattice realizations.
This difference is due to the contribution of the {}``quasi-divergent''
energy modes that emerge when $N_{a}\neq N_{b}$. We will get back
to this point in Sect.~\ref{sec:results}, presenting numerical evidence
for what we have just analysed.

%%%%%%%%%%%%%%%%%%%%%%%%%%%%%%%%%%%%%%%%%%%%%%%%%%%%%%%%%%%%%%%%%%%%%%%%%%%%%%%
%%%%%%%%%%%%%%%%%%%%%%%%%%%%%%%%%%%%%%%%%%%%%%%%%%%%%%%%%%%%%%%%%%%%%%%%%%%%%%%
\subsection{Cluster formation and periodic boundary conditions}

The study of diluted lattices requires the concept of \emph{largest
cluster}, and therefore some care is required in constructing the
effective lattice where the quantum problem is to be solved. Since
we are interested in dilution, the algorithms discussed in Subsecs.
\ref{Cholesky} and \ref{sub:BV} are to be implemented not on all
occupied lattice sites, but only on the sites defined by the largest
connected cluster of spins, since in the thermodynamic limit a finite
magnetization cannot exist if one is below the percolation critical
threshold $p_{\text{c}}$. The dilution is induced in the lattice by diluting
any site with probability $1-p$. For $p=1$ there is no dilution
at all. When $p=p_{\text{c}}$ a classical percolation transition occurs
in the thermodynamic limit preventing the existence of magnetic long
range order in the system. According to Suding and Ziff,\cite{SZ99}
$p_{\text{c}}=0.697043(3)$ in the honeycomb lattice. Here we use
$p_{\text{c}}=0.697043$.

\begin{figure}
\begin{center}\includegraphics[
  width=8cm]{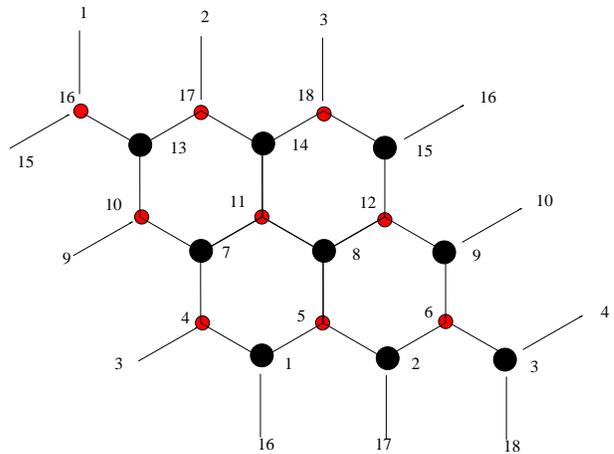}\end{center}

\caption{(color online) A finite size honeycomb lattice showing the periodic
boundary conditions used in the numerical calculations.}

\label{cap:honey}
\end{figure}

We work with finite size lattices where periodic boundary conditions
(p.b.c.) are implemented as defined in Fig.~\ref{cap:honey}. In 
Fig.~\ref{cap:honey} the links on the border are labeled according to which
site they connect to. The lattices are characterized
by their linear dimension $L$ ($L=3$ in Fig.~\ref{cap:honey}). The
total number of sites for a given $L$ is $2L^{2}$.%

\begin{figure}
\begin{center}\includegraphics[%
  scale=0.36]{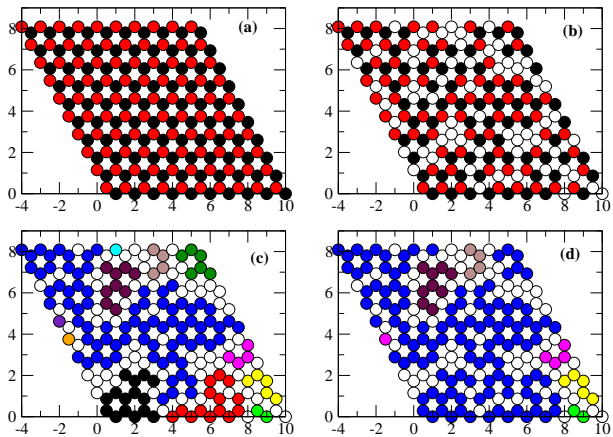}\end{center}

\caption{(color online) An example, for a disorder realization in a lattice
with $L=10$, of the cluster formation. In \textbf{(a)} is the original
lattice; in \textbf{(b)} each site is chosen to be diluted with probability
$1-p_{\text{c}}$ with $p_{\text{c}}=0.697043$; in \textbf{(c)} the several (12
in this case) clusters with rigid boundary conditions have been determined;
in \textbf{(d)} the larger cluster found in \textbf{(c)} is augmented
by the periodic boundary conditions.}

\label{cap:perc}
\end{figure}

The algorithm starts by identifying the largest cluster, for rigid
boundary conditions (this is, with no p.b.c.). As in Ref.~\onlinecite{MNC04}, 
it is only after the largest cluster is found that we apply p.b.c. to 
the original lattice, checking whether there are new sites belonging to
the largest cluster. As previously discussed in Subsect.~\ref{sub:Zm}, 
only clusters with $N_{a}=N_{b}$ are to be used, so we reject all disordered
lattice realizations in which $N_{a}\neq N_{b}$.%
\footnote{This procedure is a highly inefficient one, 
because only a small fraction ($\sim6$\% for $L=14$) of disordered lattice
realization are accepted. Nevertheless, the amount of time spent finding 
clusters with $N_{a}=N_{b}$ is a small percentage ($\sim15$\% and $\sim6$\% 
for $L=14$ in the \emph{Cholesky decomposition method} and 
\emph{Bogoliubov-Valatin transformation method}, respectively) of the time 
consumed by the diagonalization subroutines.%
} Finally, the eigenvalue problem is solved for the final cluster using
the aforementioned algorithms. In Fig.~\ref{cap:perc} we show an
example of a disorder realization and the corresponding cluster labeling
process at $p=p_{\text{c}}$. The larger cluster found for rigid boundary
conditions can be seen in panel (\textbf{c}) of Fig.~\ref{cap:perc}.
After p.b.c. implementation the final cluster has a larger number
of elements (panel (\textbf{d}) in Fig.~\ref{cap:perc}).%

%%%%%%%%%%%%%%%%%%%%%%%%%%%%%%%%%%%%%%%%%%%%%%%%%%%%%%%%%%%%%%%%%%%%%%%%%%%%%%%
%%%%%%%%%%%%%%%%%%%%%%%%%%%%%%%%%%%%%%%%%%%%%%%%%%%%%%%%%%%%%%%%%%%%%%%%%%%%%%%
\subsection{Finite-size scaling}
\label{sub:Finite-size-scaling}

The eigenvalue problem determines all the eigenvalues and eigenfunctions
for the cluster, and from these the corrections to the staggered magnetization
are computed according to Eq.~\eqref{eq:deltamz}. For a given $p$
value, $N_{\text{rz}}$ disordered lattice realizations with $N_{a}=N_{b}$
are performed, leading to an average staggered magnetization density
$m_{\text{av}}$
\begin{equation}
\label{eq:mav}
m_{\text{av}}(p,L) = \frac{1}{N_{\text{rz}}}
 \sum_{i=1}^{N_{\text{rz}}}\frac{M_{z}^{\text{stagg,}i}}{N_{m}^{i}}\,,
\end{equation}
where $M_{z}^{\text{stagg,}i}$ is the value of Eq.~\eqref{eq:magz},
and $N_{m}^{i}$ is the total number of magnetic (undiluted) sites
in the lattice, for the given disorder realization $i$. Although
$m_{\text{av}}$ does not depend explicitly on $L$, the sizes of the clusters
are determined by $L$, and therefore different $L$'s lead to different
values for Eq.~\eqref{eq:mav}. With this definition we will be able to
identify $m_{\text{av}}(p,L\rightarrow\infty)$ with the ordered magnetic
moment magnitude per magnetic ion measured in neutron diffraction
experiments.

From Eq.~\eqref{eq:magz} it is easily seen that $m_{\text{av}}$ can be
expressed as the average product of two different contributions, one
purely classical ($m_{\text{cl}}^{i}$) and the other purely quantum 
($m_{\text{qm}}^{i}$),
\begin{equation}
m_{\text{av}}(p,L)=\frac{1}{N_{\text{rz}}}\sum_{i=1}^{N_{\text{rz}}}
 m_{\text{cl}}^{i}m_{\text{qm}}^{i}\,,
\label{eq:mavMclMqm}
\end{equation}
where we used the notation $m_{\text{cl}}^{i}=
\frac{N_{\text{c}}^{i}}{N_{m}^{i}}$ for the classical factor, 
with $N_{\text{c}}^{i}=N_{a}^{i}+N_{b}^{i}$, and 
$m_{\text{qm}}^{i}=S-\delta m_{z}^{i}$ for the quantum mechanical factor.
The quantum contribution is simply the staggered magnetization density
of the larger cluster found in the disorder realization $i$. It would
be $S$ in the N\'{e}el state but it is reduced by $\delta m_{z}^{i}(p,L)$
due to quantum fluctuations, whose strength depend on dilution $p$
and lattice size $L$. If LRO is present we can assume that the sublattice
magnetization, or equivalently the staggered magnetization, is a self-averaging
quantity, as was shown to happen in the square lattice case.\cite{San02}
Thus, in the thermodynamic limit of $m_{\text{av}}$ each disorder realization
$m_{\text{qm}}^{i}$ can then be replaced by its infinite-size-extrapolated
average, which we denote by $m_{\text{qm}}$,
\begin{equation}
m_{\text{av}}(p,L\rightarrow\infty)=m_{\text{cl}}m_{\text{qm}}\,.
\label{eq:mavLinfty}
\end{equation}
The classical factor now assumes the standard form for the order
parameter of the classical percolation problem,
\begin{equation}
m_{\text{cl}}=
 \left\langle \frac{N_{\text{c}}}{N_{m}}\right\rangle _{L\rightarrow\infty}\,,
\label{eq:mcl}
\end{equation}
which is zero for $p\leq p_{\text{c}}$. Therefore a quantum critical point
can only exist above $p_{\text{c}}$ if $m_{\text{qm}}=0$ for some 
$p^{*}>p_{\text{c}}$.
To find $m_{\text{qm}}$ we need to compute the average infinite-size value
of the quantum corrections $\delta m_{z}^{\infty}$ from our finite
size calculations. We show that finite-size scaling can be found for
this quantity, from which results holding in the thermodynamic limit
can be obtained. In our study the size of the largest connected cluster
$N_{a}+N_{b}$ is not fixed, instead the linear dimension of the lattice
$L$ is. As shown for the square lattice,\cite{San02} the alternative
approach where the percolating cluster size is fixed leads to the
same magnetization value in the thermodynamic limit. The finite-size 
scaling properties of the quantum correction to the magnetization are
strictly not known for a disordered system at the percolation point.
However, in practice a direct generalization of the pure-system
scaling, using the fractal (Hausdorff) dimensionality, has been 
shown to work well.\cite{San02} Hence we will assume
\begin{equation} \label{eq:scaling}
\left\langle \delta m_{z}(p,L)\right\rangle _{N_{\text{rz}}}=
 \delta m_{z}^{\infty}+aL^{-D/2}+bL^{-D}\,,
\end{equation}
where $\delta m_{z}^{\infty}$ is the average quantum correction
to the staggered magnetization density in the thermodynamic limit,
and $D$ is the fractal dimension of the cluster, which should have
the universal value $D=91/48$ at $p_{\text{c}}$ (in two dimensions),
as is confirmed for the square and triangular lattices.\cite{PS90}

%%%%%%%%%%%%%%%%%%%%%%%%%%%%%%%%%%%%%%%%%%%%%%%%%%%%%%%%%%%%%%%%%%%%%%%%%%%%%%%
%%%%%%%%%%%%%%%%%%%%%%%%%%%%%%%%%%%%%%%%%%%%%%%%%%%%%%%%%%%%%%%%%%%%%%%%%%%%%%%
\subsection{Density of states}
\label{sub:DOSrm}

The real space diagonalization procedure, either \emph{Bogoliubov-Valatin}
or \emph{Cholesky decomposition}, is very time consuming, preventing
us from accessing large clusters (in the honeycomb lattice $L=16$
is our upper limit). Although for the staggered magnetization density
a finite-size scaling analysis can be done, we cannot easily guess
the thermodynamic limit behaviour of the DOS from results of systems
as small as $L=16$.

In this work the well known recursion method is used to compute the
average DOS. With this method we can handle lattices as large as $L=128$,
with the advantage that the obtained DOS is not the typical finite
size DOS of a system with $L=128$, but instead a very good approximation
for its thermodynamic limit value, guessed from this finite-size system.
We refer the reader to the paper of R. Haydock\cite{Hayd80} for details
in the case of non-interacting fermionic systems. Being a real space
method the effect of disorder can be easily incorporated. Here we
adopt the formulation introduced in Ref.~\onlinecite{HHT95} for disordered
electronic systems. Further details on the recursion method in relation
to disordered bosonic bilinear systems [such as model Hamiltonian
\eqref{eq:Hsw}] will be presented elsewhere.\cite{Castro05}

It is worth mentioning that the recursion method has proved to be
a powerful technique even in the presence of interactions.\cite{Hay00}
Actually, the continued fraction representation of the Fourier components
of the one particle propagator, the basis of the recursion method,
is also an essential point in the Pad\'{e} analytical continuation which
usually arises in the many-body problem.\cite{BGM00}

We define the following set of zero temperature retarded Green's functions
in the standard way,
\begin{equation} \label{eq:GRt}
\begin{split}
G_{ij}^{ab}(t) & = -i\left\langle 0\right|\{ a_{i}^{\dagger}(t),
 b_{j}^{\dagger}(0)\}\left|0\right\rangle \Theta(t),\\
G_{ij}^{ba}(t) & = -i\left\langle 0\right|\{ b_{i}(t),a_{j}(0)\}
 \left|0\right\rangle \Theta(t),\\
G_{ij}^{aa}(t) & = -i\left\langle 0\right|\{ a_{i}^{\dagger}(t),
 a_{j}(0)\}\left|0\right\rangle \Theta(t),\\
G_{ij}^{bb}(t) & = -i\left\langle 0\right|\{ b_{i}(t),b_{j}^{\dagger}(0)\}
 \left|0\right\rangle \Theta(t),
\end{split}
\end{equation}
where the notation $\left|0\right\rangle $ is used for the ground
state of the spin wave Hamiltonian \eqref{eq:Hsw}. The Fourier components
of each of the Green's functions in Eq.~\eqref{eq:GRt},
\begin{equation}
G_{ij}(E+i0^{+}) =
 \int_{-\infty}^{\infty}dt\, e^{i(E+i0^{+})t}G_{ij}(t),
\label{eq:GRe}
\end{equation}
are the quantities of interest when determining the DOS. Defining
the DOS as
\begin{equation}
\rho(E)=\frac{1}{N_{\text{c}}}\Biggl[\sum_{n=1}^{N_{a}}
 \delta(E-\omega_{n}^{(\alpha)})
 +\sum_{n=1}^{N_{b}}\delta(E-\omega_{n}^{(\beta)})\Biggr],
\label{eq:DOS}
\end{equation}
it can be easily shown that $\rho(E)$ is given in terms of the Fourier
components of the Green's functions \eqref{eq:GRe} as
\begin{multline} \label{eq:DOSG}
\rho(E) = -\frac{1}{N_{\text{c}}}\frac{1}{\pi}
 \text{Im}\Biggl[\sum_{i\in A}G_{ii}^{aa}(E+i0^{+})\\
- \sum_{i\in B}G_{ii}^{bb}(E+i0^{+}) - \sum_{i\in A}G_{ii}^{aa}(-E+i0^{+})\\
 +\sum_{i\in B}G_{ii}^{bb}(-E+i0^{+})\Biggr].
 \end{multline}
The recursion method gives $\text{Im}[G_{ij}(E+i0^{+})]$ directly,
the imaginary part of the Fourier components defined in 
Eq.~\eqref{eq:GRe},\cite{Castro05} from which the DOS is straightforwardly 
computed.

%%%%%%%%%%%%%%%%%%%%%%%%%%%%%%%%%%%%%%%%%%%%%%%%%%%%%%%%%%%%%%%%%%%%%%%%%%%%%%%
%%%%%%%%%%%%%%%%%%%%%%%%%%%%%%%%%%%%%%%%%%%%%%%%%%%%%%%%%%%%%%%%%%%%%%%%%%%%%%%
%%%%%%%%%%%%%%%%%%%%%%%%%%%%%%%%%%%%%%%%%%%%%%%%%%%%%%%%%%%%%%%%%%%%%%%%%%%%%%%
\section{Results}
\label{sec:results}

%%%%%%%%%%%%%%%%%%%%%%%%%%%%%%%%%%%%%%%%%%%%%%%%%%%%%%%%%%%%%%%%%%%%%%%%%%%%%%%
%%%%%%%%%%%%%%%%%%%%%%%%%%%%%%%%%%%%%%%%%%%%%%%%%%%%%%%%%%%%%%%%%%%%%%%%%%%%%%%
\subsection{Larger cluster statistics}
\label{subsec:clusterstat}

The number of sites in a regular planar lattice goes as the square
of its linear size. In the thermodynamic limit, the same scaling applies
to the largest cluster of the corresponding randomly-site-diluted
lattice. This behaviour persists up to the percolation threshold,
at which point the lattice is dominated by a spanning cluster of \emph{fractal}
dimension. Beyond percolation, individual clusters are no longer extensive:
they each constitute a vanishing fraction of the total number of sites.

\begin{figure}
\begin{center}\includegraphics[%
  scale=0.85]{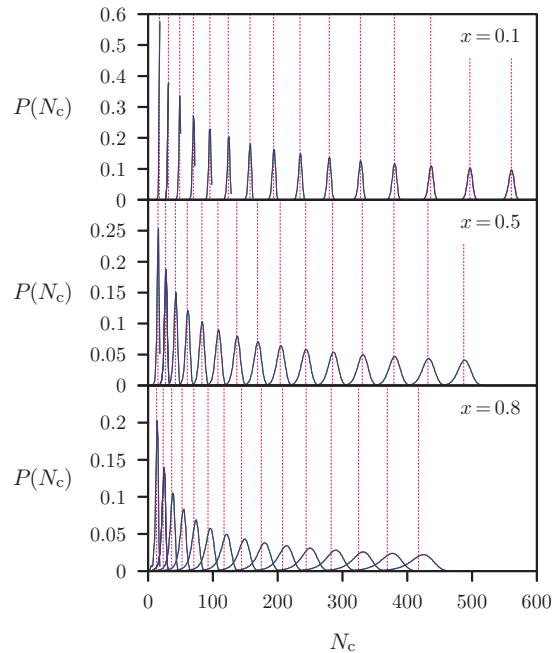}\end{center}

\caption{\label{cap:FIG:clustersize} (color online) The solid (blue) lines
show the distribution of the number of sites in the largest cluster
of a randomly site-diluted honeycomb lattice. From top to bottom,
the three panels correspond to dilution levels $x=0.1,0.5,0.8$. From
left to right, the peaks correspond to linear system sizes $L=4,5,\ldots,18$.
The (red) vertical dotted lines indicated the average cluster sizes
$\bar{N}_{\text{c}}$, computed as per Eq.~\eqref{EQ:Nclave}.}
\end{figure}

For a honeycomb lattice of size $L$ and dilution level 
$x=(1-p)/(1-p_{\text{c}})$, let $P(N_{\text{c}}|L,x)$ denote the probability 
that the largest cluster has $N_{\text{c}}$ sites. The average size of 
the largest cluster is simply the corresponding first moment:
\begin{equation}
\bar{N}_{\text{c}}(L,x)=\sum_{N_{\text{c}}=1}^{2L^{2}}N_{\text{c}}
P(N_{\text{c}}|L,x).
\label{EQ:Nclave}
\end{equation}
Example probability distributions for the honeycomb lattice are given
in Fig.~\ref{cap:FIG:clustersize}. For small $x$, the distributions
are sharply peaked. As $x\rightarrow1$, they become progressively
broader and develop long tails skewed toward small values of $N_{\text{c}}$
(marking the evolution to a different universal scaling function at
percolation).%

\begin{figure}
\begin{center}\includegraphics[%
  scale=0.85]{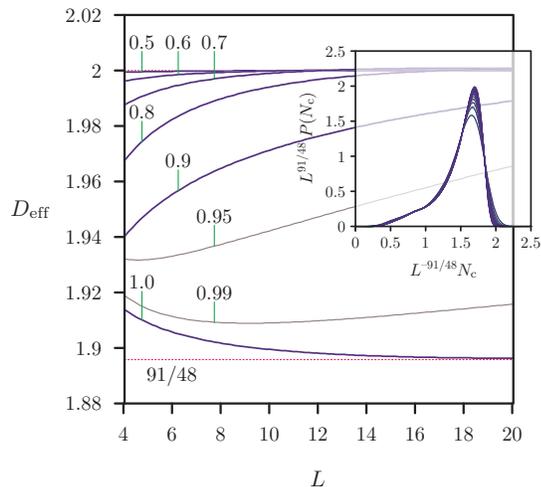}\end{center}

\caption{\label{cap:FIG:clusterscaling} (color online) The effective scaling
dimension of the largest cluster takes one of two values in the 
$L\rightarrow\infty$ limit: $D_{\text{eff}}=2$ ($0\leq x<1$) or 
$D_{\text{eff}}=91/48$ ($x=1$). For $x\lesssim0.5$, $D_{\text{eff}}$ is close 
to its asymptotic value at all systems sizes. When $x$ is close to 1, 
very large system sizes are necessary to reach the asymptotic regime. 
The figure inset shows the largest-cluster size distribution at percolation 
plotted in reduced coordinates. Each curve is computed as a histogram over
$10^{5}$ disorder realizations for system sizes $L=5,6,\ldots,48$.
As $L\rightarrow\infty$, the finite-size results converge to a smooth
scaling function (one not dissimilar from that of the square-lattice
case; see Fig.~2 of Ref.~\onlinecite{San02}).}
\end{figure}

An effective scaling dimension $D_{\text{eff}}(L,x)$ can be defined
by the relation $\bar{N}_{\text{c}}\sim L^{D_{\text{eff}}}$. Its
evolution with $L$ is plotted in Fig.~\ref{cap:FIG:clusterscaling}.
Note that $D_{\text{eff}}(L,x)$ has two points of attraction in the
limit $L\rightarrow\infty$: $D_{\text{eff}}(L,x<1)\rightarrow2$
and $D_{\text{eff}}(L,1)\rightarrow91/48$. Plotted in the appropriate
reduced coordinates---\emph{viz}., $L^{D}P(N_{\text{c}})$ versus
$L^{-D}N_{\text{c}}$ where $D=2$ below percolation and $D=91/48$
at percolation---the probability distribution tends to either a simple
delta function or the nontrivial curve shown in the inset of 
Fig.~\ref{cap:FIG:clusterscaling}.%

As can be seen in Fig.~\ref{cap:FIG:clusterscaling} (inset), a long
tail is present for smaller cluster sizes. This enhancement of the
larger cluster size distribution can be understood as a consequence
of the many possible disorder configurations for the same dilution.
That is, we can have various smaller clusters instead of one large
dominant cluster for the same number of diluted sites, though, of
course, these disorder configurations are not so favorable.

%%%%%%%%%%%%%%%%%%%%%%%%%%%%%%%%%%%%%%%%%%%%%%%%%%%%%%%%%%%%%%%%%%%%%%%%%%%%%%%
%%%%%%%%%%%%%%%%%%%%%%%%%%%%%%%%%%%%%%%%%%%%%%%%%%%%%%%%%%%%%%%%%%%%%%%%%%%%%%%
\subsection{Finite size scaling analysis}

We have performed numerical real space diagonalization of model Hamiltonian
\eqref{eq:Hsw}, as described in Sect.~\ref{sec:numeric}, for the
honeycomb and the square lattices. Lattices with sizes $L=5,6,7,8,9,10,11,12,
13,14,15,16$ (honeycomb) and $L=6,8,10,12,14,16,18,20,22,24,26,28$ (square) 
were generated. Averages were taken over $N_{\text{rz}}=10^{5}$ disorder 
realizations.%
\footnote{As the simulated lattices have $N=2\times L\times L$ sites, and the
dilution is achieved generating a random number $r\in[0,1]$ at each
lattice site, we need to be careful with the period of the random
number generator when averaging over $10^{5}$ disorder realizations.
In this work we have used the maximally equidistributed combined Tausworthe
generator,\cite{LEc96} as implemented in the \emph{GNU Scientific
Library}. The period of this generator is $2^{88}$ ($\sim10^{26}$). %
}

\begin{figure}
\begin{center}\includegraphics[%
  scale=0.75]{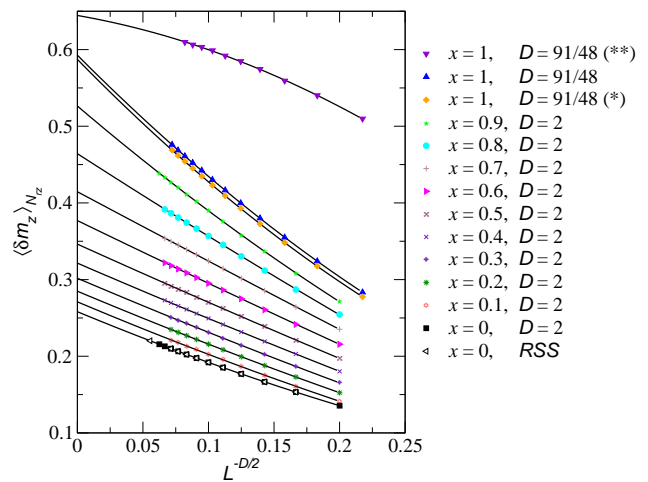}\end{center}

\caption{\label{cap:fssmz}(color online) Finite size scaling of 
$\left\langle \delta m_{z}\right\rangle $ for different values of 
$x=(1-p)/(1-p_{\text{c}})$, obtained after $10^{5}$ disorder realizations 
of lattices with equal number of sites in each sublattice. Also shown for 
$x=1$ is the result obtained when the realized lattices are not constrained 
to have $N_{a}=N_{b}$: ({*}) zero modes were subtracted and the highest 
amplitude (nonzero) mode (see text) was subtracted if $N_{a}\neq N_{b}$;
({*}{*}) only zero modes were subtracted. For $x=0$ the \emph{RSS} result 
was obtained by a \emph{reciprocal space sum} using the analytical 
result.\cite{PAB04}}
\end{figure}

In Figure \ref{cap:fssmz} we show, for the honeycomb lattice, the
average quantum correction to the staggered magnetization $\left\langle 
\delta m_{z}(p,L)\right\rangle _{N_{\text{rz}}}$,
for various values of dilution $x=(1-p)/(1-p_{\text{c}})$, as a function
of lattice size $L^{-D/2}$. The error bars are much smaller than
the symbols used. The lines are fits to the points using the finite-size
scaling hypotheses \eqref{eq:scaling}. The extrapolated zero abscissa
value gives the average quantum correction to the staggered magnetization
density in the thermodynamic limit $\delta m_{z}^{\infty}(p)$. In
the undiluted case there is an excellent agreement between the real
space diagonalization results (left-triangles) and the reciprocal
space sum (black squares), obtained from the first $\mathbf{k}$-summation
in Eq.~(C9) of Ref.~\onlinecite{PAB04}, thus providing a reliability
test to our algorithms.%

For $p=p_{\text{c}}$ we show in Fig.~\ref{cap:fssmz} the results obtained
from three different approaches. The blue up-triangles are the results
of our standard technique discussed in Sect.~\ref{sec:numeric}, \emph{i.e.},
only lattices in which $N_{a}=N_{b}$ were considered and zero modes
subtracted. The result labeled by violet down-triangles refers to
a calculation in which the disordered realized lattices are not constrained
to have $N_{a}=N_{b}$. The considerable difference between these two
results is due to the presence of one {}``quasi-divergent'' low
energy (nonzero) mode when $N_{a}\neq N_{b}$. That is, even though
we subtract the zero energy Goldstone mode as discussed in 
Sect.~\ref{sec:numeric} for $N_{a}\neq N_{b}$, there is, in this situation, 
a low energy eigenstate that contributes in order $O(1)$ for $\delta m_{z}$,
compared to the $O(1/N_{\text{c}})$ contributions of the others eigenstates.
If the contribution of this mode is subtracted the result labeled
by orange diamonds is obtained, which agrees well with the result
of our standard technique (where the constrain $N_{a}=N_{b}$ is always
used).

\begin{figure}
\begin{center}\includegraphics[%
  clip,
  scale=0.75]{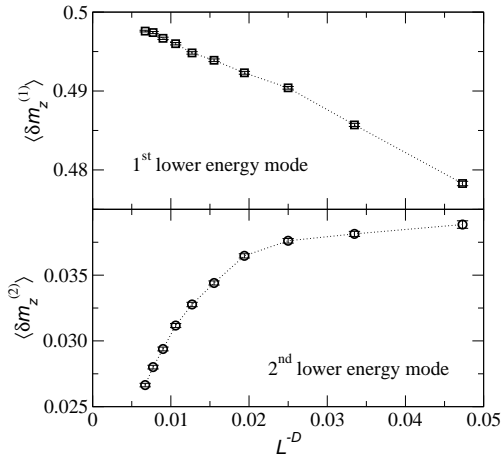}\end{center}

\caption{\label{cap:lm-honey}Contributions to $\delta m_{z}$ from: the lower
nonzero energy mode (upper panel); the lower energy mode higher than
the lower nonzero energy mode (lower panel). The average was taken
over $10^{4}$ disordered honeycomb lattices, with $N_{a}-N_{b}=\pm1$,
at $p=p_{\text{c}}$.}
\end{figure}

To better understand the presence of this nonzero energy {}``quasi-divergent''
mode when $N_{a}\neq N_{b}$, we have computed the contribution to
$\delta m_{z}$ from the lower nonzero energy mode ($\delta m_{z}^{(1)}$),
and the next one in energy ($\delta m_{z}^{(2)}$), constrained to
lattices with $N_{a}-N_{b}=\pm1$. Figure \ref{cap:lm-honey} shows
the behaviour of $\delta m_{z}^{(1)}$ (upper panel) and $\delta m_{z}^{(2)}$
(lower panel) with the average cluster size $\bar{N}_{\text{c}}\propto L^{D}$.
The $\delta m_{z}^{(2)}$ contribution decreases with $\bar{N}_{\text{c}}$,
signaling the linear increase of the number of modes that contribute
to $\delta m_{z}$. Instead, the contribution $\delta m_{z}^{(1)}$
increases with $\bar{N}_{\text{c}}$, and will be of $O(1)$ in the thermodynamic
limit. As already mentioned in Sect.~\ref{sec:numeric}, if $N_{a}$
and $N_{b}$ are both of magnitude $10^{23}$, then, if $N_{a}-N_{b}=\pm1$,
there will be, for any practical purpose, two Goldstone modes and
not only one. This statement should always be true if 
$\left|N_{a}-N_{b}\right|\ll N_{a}\sim N_{b}$. The results presented in top
panel of Fig.~\ref{cap:lm-honey} agree with this general picture. 
Furthermore, they imply that even for small sizes there is a mode, which will 
be identified with a Goldstone mode in the thermodynamic limit, 
that contributes {}``macroscopically'' to $\delta m_{z}$, though having a 
finite energy.%

%%%%%%%%%%%%%%%%%%%%%%%%%%%%%%%%%%%%%%%%%%%%%%%%%%%%%%%%%%%%%%%%%%%%%%%%%%%%%%%
%%%%%%%%%%%%%%%%%%%%%%%%%%%%%%%%%%%%%%%%%%%%%%%%%%%%%%%%%%%%%%%%%%%%%%%%%%%%%%%
\subsection{Staggered magnetization}

\begin{figure}
\begin{center}\includegraphics[%
  scale=0.75]{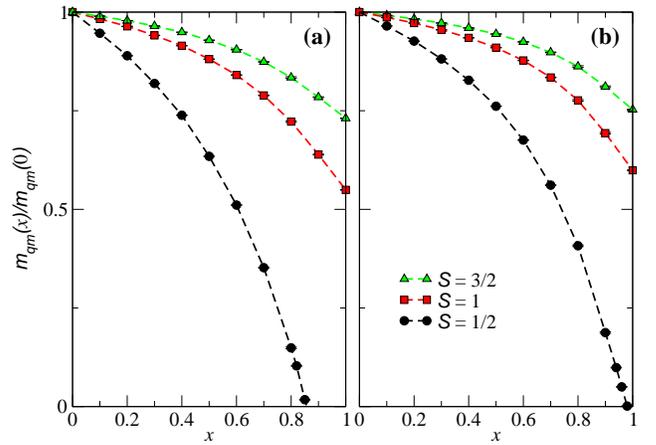}\end{center}

\caption{\label{cap:mz}(color online) Average quantum mechanical factor 
$m_{\text{qm}}(x)$ vs dilution $x=(1-p)/(1-p_{\text{c}})$ for different 
values of spin the $S$. Panel (\textbf{a}) shows the results for the 
honeycomb lattice and panel (\textbf{b}) for the square lattice.}
\end{figure}

The results we found for the quantum mechanical factor $m_{\text{qm}}(x)$
are summarized in Fig.~\ref{cap:mz} for the honeycomb lattice (panel
(\textbf{a})) and for the square lattice (panel (\textbf{b})). Three
different values of spin, $S=\frac{1}{2},1,\frac{3}{2}$, are shown.

In the undiluted limit we obtain $\delta m_{z}(0)\approx0.258$
for the honeycomb lattice, and $\delta m_{z}(0)\approx0.197$ for
the square lattice. These results are in excellent agreement with quantum
Monte Carlo results, namely, $\delta m_{z}(0)=0.2323(6)$ for the
spin $1/2$ Heisenberg antiferromagnet in the honeycomb lattice (see
Subsect.~\ref{sub:QMC}), and $\delta m_{z}(0)=0.1930(3)$ in the square
lattice.\cite{San97}

The effect of the classical factor $m_{\text{cl}}(x)$ (not shown) is only
significant very close to $p_{\text{c}}$, where it vanishes with exponent
$5/36$.\cite{SAperc} Thus, for $S>\frac{1}{2}$ there is a classically
driven order disorder transition at $p_{\text{c}}$. For $S=\frac{1}{2}$
linear spin wave theory predicts a \emph{quantum critical point} in
both the honeycomb and square lattices to occur at $x^{*}=0.85(1)$
and $x^{*}=0.98(1)$, respectively. Similar results for the square
lattice were obtained in Ref.~\onlinecite{MNC04}, though the limited
number of averages over disorder prevented the authors to distinguish
$x^{*}$ from $x=1$.

The predicted quantum critical point is absent in quantum Monte Carlo
calculations, either in the honeycomb lattice or in the square 
lattice.\cite{San02} As already mentioned in Sect.~\ref{sec:hamilt}, 
we should not expect the validity of spin wave approximation when 
$\delta m_{z}\sim S$, because inequalities \eqref{eq:ineqAb} and 
\eqref{eq:ineqBb} break down in this situation. This is precisely what 
happens when disorder increases for $S=\frac{1}{2}$.%

%%%%%%%%%%%%%%%%%%%%%%%%%%%%%%%%%%%%%%%%%%%%%%%%%%%%%%%%%%%%%%%%%%%%%%%%%%%%%%%
\subsubsection*{Comparison with experimental results}

Now we compare our results for the staggered magnetization in the
spin wave approximation with available experimental measurements on
Mn$_{p}$Zn$_{1-p}$PS$_{3}$ and Ba(Ni$_{p}$Mg$_{1-p}$)$_{2}$V$_{2}$O$_{8}$:

\paragraph{Mn$_{p}$Zn$_{1-p}$PS$_{3}$}

\begin{figure}
\begin{center}\includegraphics[%
  scale=0.75]{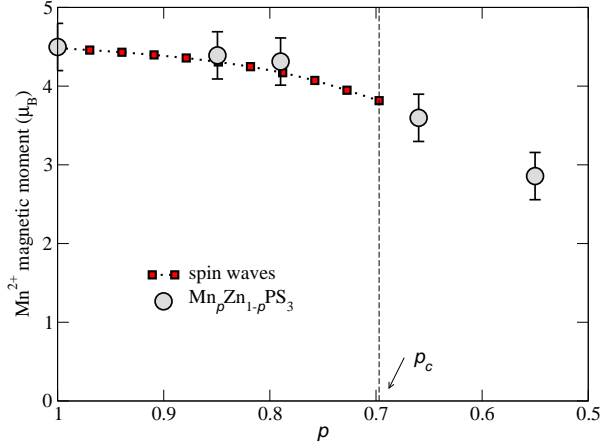}\end{center}

\caption{\label{cap:m-p.sw+exp.2S-5}(color online) Average magnetic moment
per magnetic site as function of dilution $p$. The linear spin wave
result for the $S=5/2$ Heisenberg antiferromagnet in the honeycomb
lattice (red squares) is compared with neutron scattering data on
Mn$_{p}$Zn$_{1-p}$PS$_{3}$ from Ref.~\onlinecite{GSK+00} (grey
circles).}
\end{figure}

The layered compound MnPS$_{3}$ is a $S=5/2$ Heisenberg 
antiferromagnet.\cite{KSY83} This huge spin value suggests that the spin 
wave approximation should work well in this case. Indeed, the average 
magnetic moment on the Mn atoms was found to be $4.5(2)\,\mu_{B}$ at 3.5 K 
in the pure material,\cite{GSK+00} in excellent agreement with our spin wave 
result $m\approx4.48\,\mu_{B}$. The effect of dilution in the average 
magnetic moment of Mn$^{2+}$ ions is presented in 
Fig.~\ref{cap:m-p.sw+exp.2S-5}. Neutron diffraction results on 
Mn$_{p}$Zn$_{1-p}$PS$_{3}$ are shown as grey circles,\cite{GSK+00}
and the red squares are the theoretical results within the linear
spin wave approximation. To go beyond $p_{\text{c}}$ (the first-nearest
neighbor percolation threshold) we would have to take into account
second- and third-nearest neighbor couplings in Hamiltonian \eqref{eq:hamilt1}.
Nevertheless, the effect of dilution for $p\leq p_{\text{c}}$ is already
well described by the first-nearest neighbor model. Furthermore,
the agreement between experimental and theoretical results even at
$p=p_{\text{c}}$, indicates that the primarily effect of second- and 
third-nearest neighbor interactions is classical. That is, the existence of one
largest connected cluster with a finite fraction of spins in the thermodynamic
limit is guaranteed by this couplings for $p>p_{\text{c}}$, but the quantum
correction to the staggered magnetization density is determined by
the smaller first-nearest neighbors clusters belonging to this larger
one, at least for $p\gtrsim p_{\text{c}}$. Further investigations are needed
to clarify whether this is the correct picture.\cite{CP05}%

\paragraph{Ba(Ni$_{p}$Mg$_{1-p}$)$_{2}$V$_{2}$O$_{8}$}

The layered compound BaNi$_{2}$V$_{2}$O$_{8}$ is a spin $S=1$
antiferromagnet in a honeycomb lattice. Neutron diffraction experiments
have found, in the pure case, an average magnetic moment of $1.55(4)\,\mu_{B}$
for Ni at 8 K,\cite{RHL+02} which is in good agreement with the spin
wave result $m\approx1.48\,\mu_{B}$. To our knowledge, the magnetic
moment has not yet been measured for the diluted compound. Nevertheless,
the available magnetic susceptibility measurements on 
Ba(Ni$_{p}$Mg$_{1-p}$)$_{2}$V$_{2}$O$_{8}$ for dilutions in the range 
$0.84\leq p\leq1$, show that the N\'{e}el temperature is strongly dependent 
on the amount of dilution.\cite{RHL+02} For the highest diluted sample 
($p=0.84$) a reduction of almost 70\% relative to the undiluted N\'{e}el 
temperature was found. It would be interesting to know whether the suppression 
of antiferromagnetic LRO by nonmagnetic impurities will occur at the classical 
percolation transition $p_{\text{c}}\simeq0.7$, as predicted in our 
calculations.

%%%%%%%%%%%%%%%%%%%%%%%%%%%%%%%%%%%%%%%%%%%%%%%%%%%%%%%%%%%%%%%%%%%%%%%%%%%%%%%
%%%%%%%%%%%%%%%%%%%%%%%%%%%%%%%%%%%%%%%%%%%%%%%%%%%%%%%%%%%%%%%%%%%%%%%%%%%%%%%
\subsection{N\'{e}el Temperature}

The N\'{e}el temperature of both Mn$_{p}$Zn$_{1-p}$PS$_{3}$ and 
Ba(Ni$_{p}$Mg$_{1-p}$)$_{2}$V$_{2}$O$_{8}$ shows a linear suppression 
with increasing dilution $1-p$,\cite{GSK+00,RHL+02}
a feature that is also seen in (quasi-2D) diluted Heisenberg antiferromagnets
with square lattice.\cite{CRS97,HKP+99,TFH+00}

Within the linear spin wave theory developed in Secs. \ref{sec:hamilt}
and \ref{sec:numeric} for diluted antiferromagnetic systems the finite
temperature staggered magnetization is given by
\begin{equation} \label{eq:magzT}
\begin{split}
M_{z}^{\text{stagg}}(T) & = 
 \left\langle \sum_{i\in A}S_{i}^{a,z}-\sum_{i\in B}S_{i}^{b,z}\right\rangle \\
 = & N_{\text{c}}\bigl(S-\delta m_{z}-\delta m_{z}^{T}(T)\bigr),
\end{split}
\end{equation}
where $\delta m_{z}$ is the zero-temperature correction to the staggered
magnetization defined in Eq.~\eqref{eq:deltamz}, and $\delta m_{z}^{T}(T)$
is the thermal correction
\begin{equation}
\delta m_{z}^{T}(T)=\sum_{n=1}^{N_{a}}\delta m_{z}^{(n,\alpha)}
 n_{B}(\omega_{n}^{(\alpha)})+\sum_{n=1}^{N_{b}}\delta m_{z}^{(n,\beta)}
 n_{B}(\omega_{n}^{(\beta)}),
\label{eq:deltamzT}
\end{equation}
with generalized $\delta m_{z}^{(n,\alpha)}$ and $\delta m_{z}^{(n,\beta)}$,
\begin{align}
\delta m_{z}^{(n,\alpha)} & = \frac{1}{N_{\text{c}}}
 \Biggl(\sum_{i\in A}|u_{ni}|^{2}+\sum_{i\in B}|v_{ni}|^{2}\Biggr),
 \label{eq:deltamzalphT}\\
\delta m_{z}^{(n,\beta)} & = \frac{1}{N_{\text{c}}}
 \Biggl(\sum_{i\in A}|w_{ni}|^{2}+\sum_{i\in B}|x_{ni}|^{2}\Biggr),
\label{eq:deltamzbetT}
\end{align}
and $n_{B}(\omega)=(e^{\omega/k_{B}T}-1)^{-1}$ is the Bose distribution
function. In the thermodynamic limit the averaged over disorder staggered
magnetization density can be expressed as
\begin{equation}
m_{\text{av}}(p,T,L\rightarrow\infty) = m_{\text{cl}}m_{\text{qm}}(T),
\label{eq:mavLinftyT}
\end{equation}
where $m_{\text{cl}}$ is the classical factor defined in Eq.~\eqref{eq:mcl},
and $m_{\text{qm}}(T)$ is the temperature dependent quantum mechanical factor,
\begin{equation}
m_{\text{qm}}(T) =
 S-\delta m_{z}^{\infty}-\delta m_{z}^{T,L\rightarrow\infty}(T).
\label{eq:mqmT}
\end{equation}

In the undiluted case the thermal correction $\delta m_{z}^{T}(p=1,T)$
can be expressed as
\begin{equation}
\delta m_{z}^{T,L}(p=1,T)=\frac{1}{N_{a}+N_{b}}\sum_{\mathbf{k}}
 \frac{h_{a}}{\sqrt{h_{a}^{2}-\left|\phi_{\mathbf{k}}\right|^{2}}}
 n_{B}(\omega_{\mathbf{k}}),
\label{eq:deltamzTnd}
\end{equation}
with $\omega_{\mathbf{k}}$ as in Eq.~\eqref{eq:Ek}, and $\phi_{\mathbf{k}}$
given by Eq.~\eqref{eq:phik}. The summation in $\mathbf{k}$ is done
in the first Brillouin zone of sublattice $A$ or $B$, and can be
replaced by an integration when $L\rightarrow\infty$. When $h_{a}=1$
the spin wave dispersion behaves as $\omega_{\mathbf{k}}\propto k$
in the long wave length limit, similarly to the square lattice case.
As a consequence the thermal correction to the staggered magnetization
develops a logarithmic divergence, which signals the well known suppression
of LRO at $T>0$ in the 2D isotropic Heisenberg model. 

Therefore, if LRO is present up to $T_{N}\neq0$, either a magnetic
anisotropy $h_{a}$ or a finite interplanar exchange $J_{\perp}$
(or both) must be present. If the former is the dominant effect $T_{N}$
can be calculated using the mean-field like equation 
\footnote{The N\'eel temperature determined from  Eq.~\eqref{eq:Tmfgap} 
for isotropic Heisenberg systems is known to
be overestimated. In order to correct for it a self consistent solution
of this equation is used, where $S$ is replaced by $m_{\rm av}$. This
procedure, very simple to implement in the undiluted case, since there 
is an analytical expression available for the magnon spectrum, is much 
more difficult in our case. We postpone the discussion of this aspect to 
a latter publication.}
\begin{equation}
m_{\text{qm}}(T_{N})=0.
\label{eq:Tmfgap}
\end{equation}
In the latter the transition should occur when the interplanar coupling
is strong enough to stabilize the LRO in comparison with thermal fluctuations:
\begin{equation}
J_{\perp}m_{\text{qm}}^{2}(p,T=0)\frac{\xi^{2}(p,T_{N})}{A/2}
\approx k_{B}T_{N},
\label{eq:TmfJperp}
\end{equation}
The parameter $\xi(p,T)$ is the inplane correlation length, which
characterizes the spin fluctuations of a layered system in a paramagnetic
phase. The area of a hexagon of side $c$ is given by $A=c^{2}3\sqrt{3}/2$.
The correlation length can be calculated in the context of the modified
spin wave theory,\cite{TAK89} and in the non-diluted ($p=1$) case
it is exponentially divergent with $1/T$ as $T\rightarrow0$. The
mean field picture which leads to Eq.~\eqref{eq:TmfJperp} was proposed
in Ref.~\onlinecite{CCN02}, and gives a good description of the variation
of $T_{N}(p)/T_{N}(0)$ with dilution $1-p$ in a variety of layered
compounds with square lattice.

In the case of MnPS$_{3}$ a small gap of magnitude $\Delta E=0.5$~meV
was found in the spin wave energy at the the Brillouin zone 
center.\cite{WRL+98} This energy gap can be explained by either a 
single-ion anisotropy or a dipole coupling, being modeled here by a small 
magnetic anisotropy $h_{a}>1$. From the spin wave dispersion \eqref{eq:Ek} 
it is found that $h_{a}\approx1.004$ is needed to obtain $\Delta E=0.5$ meV
(a nearest-neighbor exchange of magnitude $J=0.8$ meV was used\cite{WRL+98}).
We remark that such a small magnetic anisotropy has no effect in the
conclusions we have made so far based in the isotropic Heisenberg
model ($h_{a}=1$). As an example, the the average magnetic moment
on the Mn atoms given by spin-wave theory is $m\approx4.48\,\mu_{B}$
for $h_{a}=1$ and $m\approx4.55\,\mu_{B}$ for $h_{a}=1.004$, both
in excellent agreement with the experimental value $4.5(2)\,\mu_{B}$
at 3.5~K.\cite{GSK+00}  Inserting this value of $h_{a}$ into Eq.
(\ref{eq:deltamzTnd}) we obtain $T_{N}\approx70$~K as a solution
of Eq.~(\ref{eq:Tmfgap}), in agreement with the measured value 
$T_{N}=78$~K.\cite{KSY83}

Nevertheless a finite interplanar exchange of magnitude 
$J_{\perp}=0.0019(2)$~meV is also present in the MnPS$_{3}$ 
compound.\cite{WRL+98} 
With $\xi(p\!=\!1,T_{N}\!=\!78~\text{K})=27.5~\text{\AA}$
measured by neutron scattering,\cite{RWB00} and $c=3.5~\text{\AA}$,
\cite{okuda-mpolcm1986} we obtain from the mean field equation 
\eqref{eq:TmfJperp} $T_{N}\approx6$~K.
This small value of $T_{N}$ is an indication that the effect of the
interplanar coupling is not as important as the magnetic anisotropy
in stabilizing the LRO. Therefore we use Eq.~\eqref{eq:Tmfgap} to
study the effect of dilution on $T_{N}(p)$. The thermal correction 
$\delta m_{z}^{T}$ defined by Eq.~\eqref{eq:deltamzT} is computed via 
recursion method (see Subsect.~\ref{sub:DOSrm}), noting that it can be 
expressed as
\begin{equation}
\delta m_{z}^{T}(T)=\int_{0}^{\infty}dEn_{B}(E)K(E),
\label{eq:deltamzTrm}
\end{equation}
where the kernel $K(E)$ is given by
\begin{multline} \label{eq:kernel}
K(E) = -\frac{1}{N_{\text{c}}}\frac{1}{\pi}\text{Im}
 \Biggl[\sum_{i\in A}G_{ii}^{aa}(E+i0^{+})\\
+\sum_{i\in B}G_{ii}^{bb}(E+i0^{+}) + \sum_{i\in A}G_{ii}^{aa}(-E+i0^{+})\\
+\sum_{i\in B}G_{ii}^{bb}(-E+i0^{+})\Bigr].
\end{multline}
It is worth mentioning that with the recursion method $\delta m_{z}^{T}$
can be computed with the same precision (limited by the linear size
$L=128$ of the sample) from the undiluted limit $p=1$ to the percolation
threshold $p=p_{\text{c}}$.

\begin{figure}
\begin{center}\includegraphics[%
  clip,
  scale=0.75]{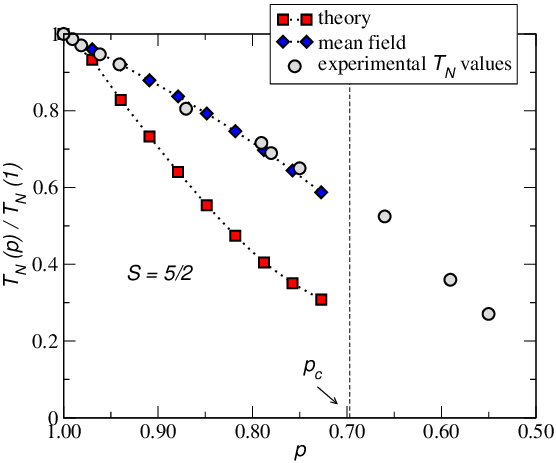}\end{center}

\caption{\label{cap:TN2S-5}(color online) $T_{N}(p)/T_{N}(0)$ vs $p$ for
$S=5/2$. Shown are the results obtained by numerically solving Eq.
\eqref{eq:Tmfgap} with $\delta m_{z}^{T}$ computed applying the
recursion method to systems with $L=128$ and averaging over 200 to
400 disorder realizations (squares), the mean-field result of Eq.
(\ref{eq:TMFp}) (diamonds), and experimental results on 
Mn$_{p}$Zn$_{1-p}$PS$_{3}$ from Ref.~\onlinecite{GSK+00} (circles).}
\end{figure}

The result of numerically solving Eq.~\eqref{eq:Tmfgap}---with 
$\delta m_{z}^{T}$ computed by applying the recursion method to systems 
with $L=128$ and averaging over 200 to 400 disorder realizations---is shown in 
Fig.~\ref{cap:TN2S-5}.%
Also shown are the results of magnetometry measurements on 
Mn$_{p}$Zn$_{1-p}$PS$_{3}$ from Ref. \onlinecite{GSK+00}. The difference 
between the theoretical results and experimental values suggests that in 
opposition to the magnetic moment at zero temperature 
(see Fig. \ref{cap:m-p.sw+exp.2S-5}) the effect of second- and 
third-nearest-neighbor couplings should be included to obtain a quantitatively 
correct N\'{e}el temperature as dilution is increased. An estimation of 
$T_{N}(p)/T_{N}(1)$ can as well be obtained by standard mean-field theory, 
$T_{N}^{MF}=\frac{2}{3}JzS(S+1)$.\cite{MJintheorPhys}
Replacing $S$ by the zero temperature staggered magnetization density
$m_{av}(p)$ defined in Eq.~(\ref{eq:mavLinfty}), and assuming that
the coordination number decreases linearly with dilution, $z\propto p$,
the ratio $T_{N}(p)/T_{N}(1)$ is given by
\begin{equation}
\frac{T_{N}(p)}{T_{N}(1)} = p\, m_{av}(p)[m_{av}(p)+1].
\label{eq:TMFp}
\end{equation}
In Fig.~\ref{cap:TN2S-5} we show as diamonds the results of Eq.
(\ref{eq:TMFp}). Although this result reproduces the correct dependence
on $p$, it should be stressed that as a mean-field approximation
the absolute value of $T_{N}(p)$ is overestimated.

The effect of dilution on the N\'{e}el temperature of 
Ba(Ni$_{p}$Mg$_{1-p}$)$_{2}$V$_{2}$O$_{8}$ was studied by Rogado 
\emph{et al.} for dilutions in the range $0.84\leq p\leq1$.\cite{RHL+02}
The few experimental results concerning the magnetic properties of
BaNi$_{2}$V$_{2}$O$_{8}$ are insufficient to undoubtedly determine
the model which better describes the magnetic behaviour of this compound.
Although electron-spin resonance measurements seem to be well fitted by a 
weakly anisotropic Heisenberg model with easy-plane symmetry (XY), \emph{i.e.},
$h_{a}\lesssim1$ in Hamiltonian \eqref{eq:haf},
the same results can as well be explained with the isotropic limit
of this model.\cite{HvN+03} Further experiments would be valuable
in determining the nature of the LRO observed in this compound, in
particular inelastic neutron scattering from which the spin wave dispersion
can be measured. Here we assume that a small gap is present at the
Brillouin zone center, and that it can be modeled by a small uniaxial
interaction anisotropy, \emph{i.e.}, $h_{a}\gtrsim1$ in Hamiltonian 
\eqref{eq:haf}. In particular $h_{a}-1\approx10^{-4}$ is needed to get 
$T_{N}\approx50$~K in the undiluted case (a nearest-neighbor exchange of 
magnitude $J\approx4$~meV was used).\cite{RHL+02}

\begin{figure}
\begin{center}\includegraphics[%
  scale=0.75]{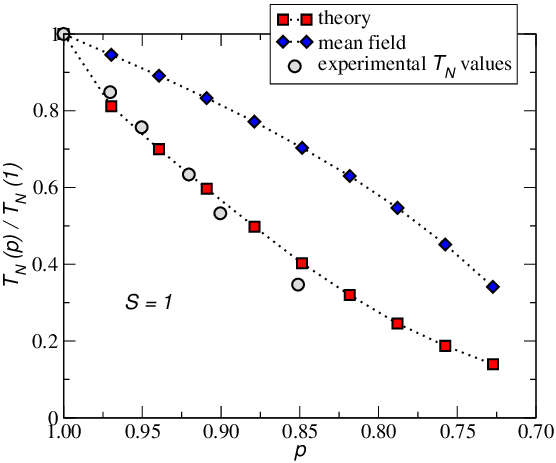}\end{center}

\caption{\label{cap:TNS-1}(color online) $T_{N}(p)/T_{N}(0)$ vs $p$ for
$S=1$. Shown are the results obtained by numerically solving 
Eq.~\eqref{eq:Tmfgap} with $\delta m_{z}^{T}$ computed applying the 
recursion method to systems with $L=128$ and averaging over 200 to 400
disorder realizations (squares), the mean-field result of Eq.~(\ref{eq:TMFp})
(diamonds), and experimental results on 
Ba(Ni$_{p}$Mg$_{1-p}$)$_{2}$V$_{2}$O$_{8}$ from Ref.~\onlinecite{RHL+02}
(circles).}
\end{figure}

The $T_{N}(p)/T_{N}(1)$ vs $p$ result obtained by numerically solving
Eq.~(\ref{eq:Tmfgap}) for $S=1$, with $\delta m_{z}^{T}$ computed
applying the recursion method to systems with $L=128$ and averaging
over 200 to 400 disorder realizations is shown in Fig.~\ref{cap:TNS-1}
(squares).%
Also shown are the mean-field result of Eq.~(\ref{eq:TMFp}) (diamonds)
and results of magnetic susceptibility data for 
Ba(Ni$_{p}$Mg$_{1-p}$)$_{2}$V$_{2}$O$_{8}$ (circles) (Ref. 
\onlinecite{RHL+02}). The disagreement between the
mean-field result (Eq.~(\ref{eq:TMFp})) and experimental values can
be attributed to the small spin $S=1$ value, which means higher quantum
fluctuations and less mean-field like behaviour. The theoretical result
(squares) and the experimental values are in reasonable agreement,
though it seems to worsen as dilution increases. It should be noted
that the spin-wave theory for layered materials is not really adequate
at $T\sim T_{N}$, and when it is applied to the mean-field like 
Eq.~(\ref{eq:Tmfgap}) it tends to overestimate the absolute value of
the N\'{e}el temperature.\cite{IKK99}

%%%%%%%%%%%%%%%%%%%%%%%%%%%%%%%%%%%%%%%%%%%%%%%%%%%%%%%%%%%%%%%%%%%%%%%%%%%%%%%
%%%%%%%%%%%%%%%%%%%%%%%%%%%%%%%%%%%%%%%%%%%%%%%%%%%%%%%%%%%%%%%%%%%%%%%%%%%%%%%
\subsection{Density of states}

The effect of dilution has a strong impact on the DOS of the system.
Since the momentum is no-longer a well defined quantum number the
spin waves acquire a finite lifetime.\cite{CCN02} As a consequence,
the basis that diagonalizes the problem has a very different energy
spectrum, which implies a different DOS.

We have calculated the DOS of the antiferromagnetic Heisenberg model
in the linear spin wave approximation for the honeycomb and square
lattices in the presence of dilution. The recursion method briefly
discussed in Subsect. \ref{sub:DOSrm} was used to study the variation
of the DOS with dilution. The method is valid from the undiluted
$p=1$ limit to the percolation threshold $p_{c}$, and enables the
access to the whole energy spectrum. The precision limit is set by
the linear size $L$ of the system, which we fix here to $L=128$
both in the honeycomb and square lattices.

\begin{figure}
\begin{center}\includegraphics[%
  clip,
  scale=0.75]{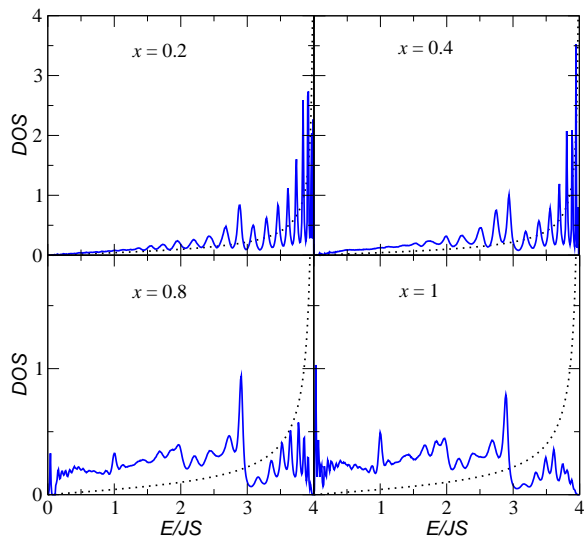}\end{center}

\caption{\label{cap:DOSsquare}(color online) DOS of the Heisenberg 
antiferromagnetic model in the linear spin wave approximation for the 
square lattice. An energy mesh with spacing $0.01$ in units of $JS$ 
was used. This results were obtained applying the recursion method to 
systems with $L=128$, and averaging over 200 to 400 disorder realizations. The
dotted line is the clean limit DOS.}
\end{figure}

In Fig.~\ref{cap:DOSsquare} we show the square lattice DOS at four
different values of dilution $x$.%
The depletion of the high energy part of the DOS in favor of low
energy modes is clearly seen as dilution is increased, in agreement
with the results obtained by exact diagonalize smaller systems.\cite{MNC04}
The two structures visible at around $E/JS=2$ and $3$, which Mucciolo
\emph{et al.}\cite{MNC04} associated with the breaking of the clean-limit
magnon branch into three distinct but broad branches, are also evident.

\begin{figure}
\begin{center}\includegraphics[%
  clip,
  scale=0.75]{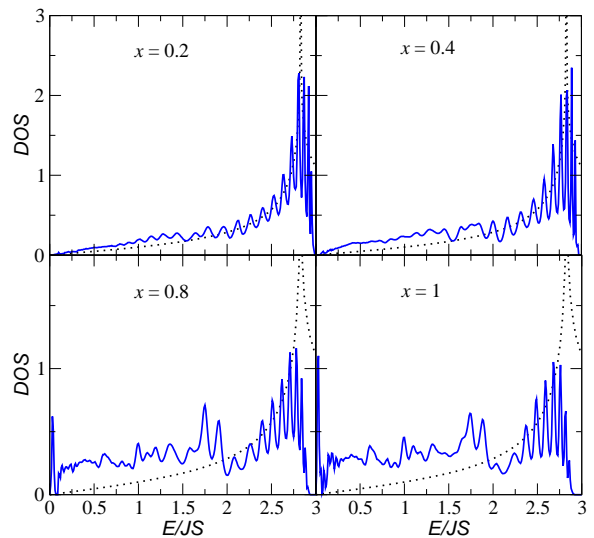}\end{center}

\caption{\label{cap:DOShoney}(color online) DOS of the Heisenberg 
antiferromagnetic model in the linear spin wave approximation for 
the honeycomb lattice. An energy mesh with spacing $0.01$ in units of 
$JS$ was used. This results were obtained applying the recursion method 
to systems with $L=128$, and averaging over 200 to 400 disorder realizations. 
The dotted line is the clean limit DOS.}
\end{figure}

The DOS for the honeycomb lattice is shown in Fig. \ref{cap:DOShoney}.
A decrease in the density of high-frequency states and the proportional
increase in the density of low-frequency ones is also clear as dilution
increases.%
This feature can then be viewed as a general effect of the presence
of dilution. Structures as those observed in the square lattice case,
just below $E/JS=2$ and 3, are not so easily identified. Nevertheless,
a feature of this kind seems to be present just below $E/JS=2$. To
determine whether or not it can be associated to the presence of fractons,
as in the square lattice case,\cite{MNC04} a more detailed study
is needed, such as the calculation of the dynamical structure factor
in the diluted honeycomb lattice.

\begin{figure}
\begin{center}\includegraphics[%
  clip,
  scale=0.75]{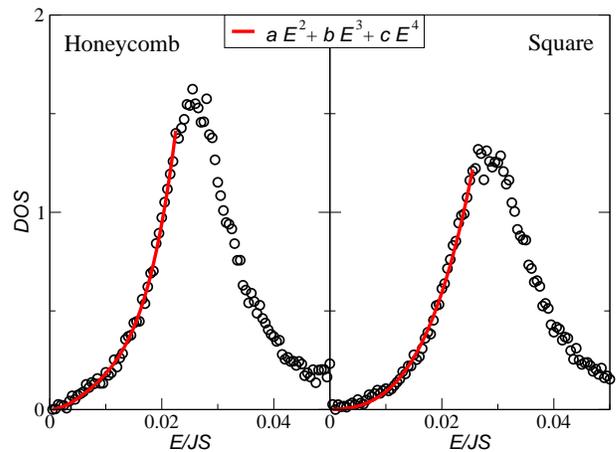}\end{center}

\caption{\label{cap:DOSlowener}(color online) Low energy behaviour of the
DOS of the Heisenberg antiferromagnetic model in the linear spin wave
approximation for the honeycomb (left) and square (right) lattices
at $x=x_{c}$. An energy mesh with spacing $5\times10^{-4}$ in units
of $JS$ was used. This results were obtained applying the recursion
method to systems with $L=128$, and averaging over 800 disorder realizations.}
\end{figure}

The effect of moving spectral weight from the top of the band to lower
energies due to dilution is accompanied by the appearance of a set
of peaks, starting to develop in the high-frequency part of the spectrum
for small dilution and extending to the entire band as dilution increases.
There is, however, a particular peak that deserves special attention.
This peak can be seen very close to the bottom of the band ($E=0$)
for $x\geq0.8$ both in the honeycomb and square lattice DOS. Figure
\ref{cap:DOSlowener} is a zoom of the DOS close to $E=0$ at $x=x_{c}$.%
Being present both in the honeycomb and square lattices, though a
bit stronger in the former, this peak seems to be a general feature
associated with dilution. In fact, it is closely related to the finiteness
of the quantum corrections to the staggered magnetization at zero
temperature.

As shown by Mucciolo \emph{et al.}\cite{MNC04}, the finiteness of
the quantum fluctuations reduces to the problem of the convergence
of the integral $\int_{0}^{E_{max}}dE\rho(E)E^{-1}$. In 
Fig.~\ref{cap:DOSlowener} we show a polynomial fit to the low-energy behaviour 
of the DOS (red line in the left side of the peak). Although it should be 
seen as guide to the eyes, we can undoubtedly say that in the low-energy limit
the DOS behaves as $\rho(E)\propto E^{\alpha}$ with $\alpha>1$,
and thus the above mentioned integral is convergent. This result is
consistent with the existence of an upper bound for the quantum fluctuations
in any model with a classically ordered ground state whose Hamiltonian
can be mapped onto that of a system of coupled harmonic oscillators,
argued by Mucciolo \emph{et al.}.\cite{MNC04} This result also agrees
with the FSS results presented in Subsect. \ref{sub:Finite-size-scaling},
where we found finite values for $\delta m_{z}^{\infty}(x)$. And
the fact that $\delta m_{z}^{\infty}(x_{c})>1/2$ can be attributed
to the bad-behaviour of the spin-wave approximation when $\delta m_{z}\sim S$,
as will be shown in the next section.

%%%%%%%%%%%%%%%%%%%%%%%%%%%%%%%%%%%%%%%%%%%%%%%%%%%%%%%%%%%%%%%%%%%%%%%%%%%%%%%
%%%%%%%%%%%%%%%%%%%%%%%%%%%%%%%%%%%%%%%%%%%%%%%%%%%%%%%%%%%%%%%%%%%%%%%%%%%%%%%
\subsection{Quantum Monte Carlo results for $S=1/2$}
\label{sub:QMC}

We have performed a Monte Carlo study of the $S=1/2$ quantum Heisenberg
antiferromagnet on the site-diluted honeycomb lattice using Stochastic
Series Expansion (SSE).~\cite{San97,San99} Unlike the spin wave
approach described in Sects.~\ref{sec:hamilt} and \ref{sec:numeric}---which
should be understood as an expansion in the relative reduction of
the staggered moment $\delta m_{z}/S$---this technique is exact (up
to statistical uncertainties) and well-behaved even when $\delta m_{z}\sim S$.
In particular, the SSE Monte Carlo can access the the small-$S$,
near-percolation regime where the spin wave calculation becomes unreliable.

We have closely followed the procedure outlined in Ref.~\onlinecite{San02},
which treats the site dilution problem on the square lattice. To accelerate
convergence, we have taken advantage of the $\beta$-doubling scheme
described therein: 100 equlibration and 200 sampling sweeps are performed
at each temperature with the resulting configuration (an $M$-element
operator list $S_{M}=[a_{1},b_{1}],\ldots,[a_{M},b_{M}]$) used to
generate a high-probability initial configuration at the next lowest
temperature ($S_{2M}=[a_{1},b_{1}],\ldots,[a_{M},b_{M}],[a_{M},b_{M}],\ldots,[a_{1},b_{1}]$)
according to the cooling schedule $\beta=2,4,8,\ldots,2048,4096$.

A refinement to previous work is that we extrapolate the staggered
magnetization to the thermodynamic limit 
using \emph{two} different quantities:
\begin{subequations} \label{EQ:minfty}
\begin{align} \label{EQ:minftyA}
m_{\text{qm}} &= \lim_{L\rightarrow\infty} \biggl\langle \frac{2}{N_{\text{c}}}\Bigl\lvert \,
\hat{M}^{\text{stagg}}_z \,\Bigr\rvert \biggr\rangle_{\!L,x},\\ \label{EQ:minftyB}
m_{\text{qm}}^2 &= \lim_{L\rightarrow\infty} \biggl\langle \frac{3}{N_{\text{c}}^2}\Bigl(
\hat{M}^{\text{stagg}}_z \Bigr)^2 \biggr\rangle_{\!L,x}.
\end{align}
\end{subequations}
Here, $\hat{M}^{\text{stagg}}_z = \sum_{i\in A} \hat{S}_i^z - \sum_{i\in B} \hat{S}_i^z$
is the $z$-projected staggered magetization and $m_{\text{qm}}$ is the quantum mechanical factor introduced
in Subsect.~\ref{sub:Finite-size-scaling}.
The notation $\langle\,\cdot\,\rangle_{L,x}$ represents an ensemble average
over the quantum states of the system and over all configurations
of the size-$L$ lattice with dilution $x$. The site indices in $\hat{M}^{\text{stagg}}_z$ are
understood to range over only the largest connected cluster.

Equation~\eqref{EQ:minftyA}, being linear, is analogous to the quantity
$S-\delta m_{z}$ computed via spin wave theory. Equation~\eqref{EQ:minftyB}
is essentially a structure factor and equivalent to Eq.~(10) of Ref.~\onlinecite{San02}.
The factors 2 and 3 in Eqs.~\eqref{EQ:minfty} are a consequence
of the rotational invariance of the ground state. Their particular
values follow from the averages $\int\! d\hat{\Omega}\,\lvert\hat{\Omega}\cdot\hat{z}\rvert=4\pi/2$
and $\int\! d\hat{\Omega}\,\bigl(\hat{\Omega}\cdot\hat{z}\bigr)^{2}=4\pi/3$
where $\hat{\Omega}$ is a vector ranging over the unit sphere. (Such
geometric factors are irrelevant to the spin wave case; there the
ground state is symmetry-broken by explicit construction.)

As in Ref.~\onlinecite{San02}, we use the straight-forward generalization 
of the finite-size scaling form for the clean system,\cite{HUS88}
\begin{subequations} \label{EQ:fssansatz}
\begin{align} 
\label{EQ:fssansatzA}
\biggl\langle \frac{2}{N_{\text{c}}}\Bigl\lvert \,
\hat{M}^{\text{stagg}}_z \,\Bigr\rvert \biggr\rangle_{\!\!L,x}^{2} & 
\!= m_{\text{qm}}^2 + \frac{a_1}{\sqrt{\bar{N}_{\text{c}}}} + \frac{a_2}{\bar{N}_{\text{c}}} + \cdots, \\ 
\label{EQ:fssansatzB}
\biggl\langle \frac{3}{N_{\text{c}}^2}\Bigl(
\hat{M}^{\text{stagg}}_z \Bigr)^2 \biggr\rangle_{\!\!L,x} &
\!= m_{\text{qm}}^2 + \frac{b_1}{\sqrt{\bar{N}_{\text{c}}}} + \frac{b_2}{\bar{N}_{\text{c}}} + \cdots.
\end{align}
\end{subequations} 
[As the discussion in Subsect.~\ref{subsec:clusterstat} makes clear, this
converges to $L^{-D/2}$ powerlaw behavior at large $L$, as in Eq.~\eqref{eq:scaling}.]
Numerical measurements of the two quantities on the left-hand side
of Eqs.~\eqref{EQ:fssansatzA} and \eqref{EQ:fssansatzB} may be
fit to the corresponding functions on the right-hand side either simulaneously---with
parameters $m_{\text{qm}}$, $\{a_{i}\}$, $\{b_{i}\}$---or separately---with parameters $m_{\text{qm}}$,
$\{a_{i}\}$ and $m_{\text{qm}}'$, $\{b_{i}\}$. Verifying that $m_{\text{qm}}\approx m_{\text{qm}}'$
serves as a consistency check.
\begin{figure}
\begin{center}\includegraphics[%
  scale=0.85]{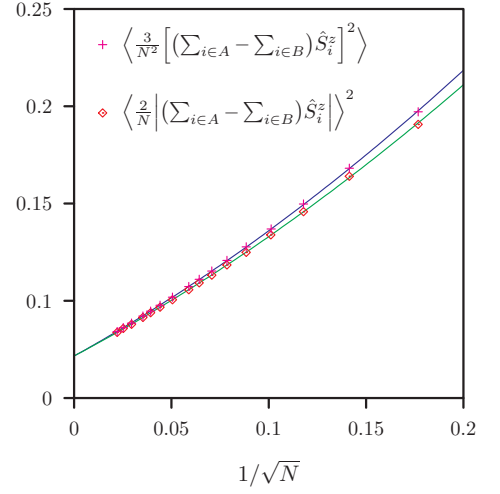}\end{center}
\caption{\label{FIG:magundiluted} The staggered magnetization of the undiluted
honeycomb lattice ($x=0$, $N_{\text{c}}=N=2L^{2}$) is extrapolated
to the thermodynamic limit following Eqs.~\eqref{EQ:fssansatzA}
and \eqref{EQ:fssansatzB}. A simultaneous fit of the two data sets
yields the value $m_{\text{av}}(L\to\infty)=0.2677(6)$.}
\end{figure}

In the case of the \emph{undiluted} honeycomb lattice
(for which $m_{\text{av}}(L\to\infty)\equiv m_{\text{qm}}$), 
we have simulated lattices up to linear size $L=32$
(\emph{i.e.}, up to $2\times32^{2}=2048$ sites). Observables were
computed using a bootstrap analysis~\cite{ET93} of 150 bins of $10^{5}$
samples each ($1.5\times10^{6}$ total Monte Carlo sweeps). Best fits
to the data, shown in Fig.~\ref{FIG:magundiluted}, give the thermodynamic
limit $m_{\text{av}}(L\rightarrow\infty)=0.2677(6)$.
This is somewhat smaller than the square lattice value $m_{\text{av}}(L\rightarrow\infty)=0.3070(3)$,\cite{San02}
a reduction that reflects the larger quantum fluctuations on the less
meanfield-like honeycomb lattice.

Note that our value of the staggered magnetization is larger than
(but consistent with) an earlier Monte Carlo measurement due to Reger
\emph{et al}.~\cite{RRY89} (within 1.6 standard deviations). It
is also, we believe, considerably more accurate. The Reger group's
value of $m_{\text{av}}(L\rightarrow\infty)=0.22(3)$
was computed by extrapolating relatively large Trotter errors ($0.1<\Delta\tau<0.2$)
to $\Delta\tau\rightarrow0$ and small systems sizes ($4<L<8$) to
$L\rightarrow\infty$. Moveover, their analysis supposes that the
inverse temperature $\beta=10$ is sufficiently cold to extract the
ground state properties of the system, which is very likely incorrect.~\cite{San02}%
\begin{figure}
\begin{center}\includegraphics[%
  scale=0.85]{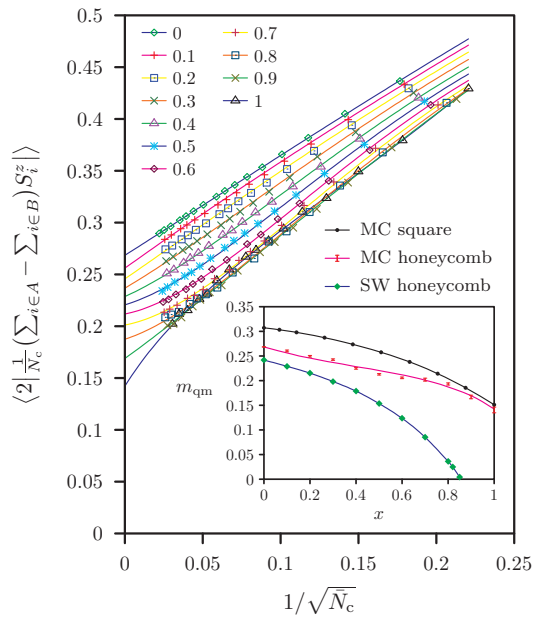}\end{center}

\caption{\label{FIG:magdiluted} (color online) The main plot shows an extrapolation
to the thermodynamic limit of twice the $z$-projected staggered magnetization
for various dilution levels $x$ (as indicated by the symbols in the
upper-left legend). The lines drawn through the data points represent
a global fit to Eqs.~\eqref{EQ:fssansatz} in which $m_{\text{qm}}(x),a_{1}(x),b_{1}(x),\ldots$
are treated as powerseries in $x$ and varied. The resulting function
$m_{\text{qm}}(x)$ appears as the solid
(pink) line in the figure inset alongside Monte Carlo results for
the square lattice (from Ref.~\onlinecite{San02}) and spinwave results
for the honeycomb lattice. The (red) errorbars indicate the values
of $m_{\text{qm}}$ extrapolated from each
fixed-$x$ dataset taken individually. }
\end{figure}

For the \emph{diluted} honeycomb lattice, we computed the staggered
magnetization as an average over $10^{5}$ randomly-generated disorder
realizations. Simulations of system sizes up to $\bar{N}_{\text{c}} \approx 2000$ were extrapolated
to the thermodynamic limit, as shown in Fig.~\ref{FIG:magdiluted}.
The figure inset illustrates the dependence of $m_{\text{qm}}$ on dilution.

In contrast to the spinwave prediction, we find that LRO
persists right up to the classical percolation threshold. The magnitude
of the staggered magnetization decreases with dilution but does not
vanish: $m_{\text{qm}}=0.139(6)$ at $x=1$, which represents a roughly $50\%$ reduction in magnetic
moment over the undiluted ($x=0$) lattice. This is comparable to
the effect seen in the square lattice where $m_{\text{qm}}(0)=0.3070(3)$
falls to $m_{\text{qm}}(1)=0.150(2)$.

\begin{figure}
\begin{center}\includegraphics[%
  scale=0.85]{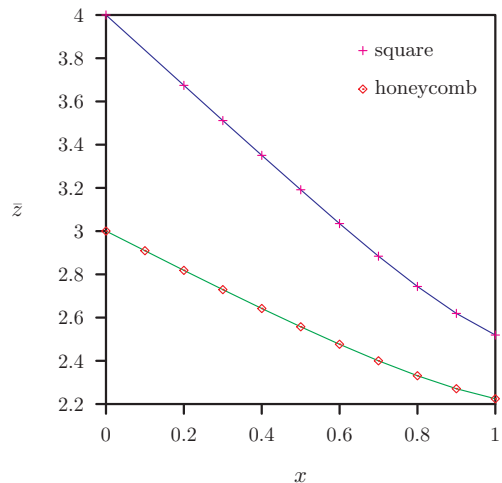}\end{center}
\caption{\label{FIG:zbar} The disorder-averaged coordination number $\bar{z}(x)$
is plotted as a function of dilution level for infinite square and
honeycomb lattices. The difference between the two lattice types narrows
as $x\rightarrow1$. The undiluted square lattice is 33\% more coordinated
than the honeycomb lattice. At percolation, it is only 12\% more so.}
\end{figure}

We observe that the square- and honeycomb-lattice values of $m_{\text{qm}}$
are remarkably close in the vicinity of $x=1$. The likely
explanation is that the percolating clusters---retaining little of
the structure of their undiluted parent lattice---are themselves quite
similar. Both have fractal dimension $D=91/48$ and a similar nearest
neighbour count: with increasing site dilution, the average coordination
number goes from $\bar{z}^{\text{hc}}(0)=3$ and $\bar{z}^{\text{sq}}(0)=4$
to $\bar{z}^{\text{hc}}(1)=2.22$ and $\bar{z}^{\text{sq}}(1)=2.52$;
see Fig.~\ref{FIG:zbar}. The Monte Carlo results are consistent
with our understanding that the quantum fluctuations disrupt the LRO 
in inverse proportion to the number of nearest neighbours contributing
to the local staggered mean field at each site.

%%%%%%%%%%%%%%%%%%%%%%%%%%%%%%%%%%%%%%%%%%%%%%%%%%%%%%%%%%%%%%%%%%%%%%%%%%%%%%%
%%%%%%%%%%%%%%%%%%%%%%%%%%%%%%%%%%%%%%%%%%%%%%%%%%%%%%%%%%%%%%%%%%%%%%%%%%%%%%%
%%%%%%%%%%%%%%%%%%%%%%%%%%%%%%%%%%%%%%%%%%%%%%%%%%%%%%%%%%%%%%%%%%%%%%%%%%%%%%%
\section{Summary and concluding remarks}
\label{sec:conclusions}

In this work we studied the magnetic properties for diluted Heisenberg
models in the honeycomb lattice. Refined results for the
density of states in the square lattice case were also reported.
We have shown that spin wave theory in diluted lattices is quite
successful in describing the magnetic properties of $S>1/2$
systems. On the other hand, for $S=1/2$, spin wave theory breaks down
and one has to approach the problem using a  Monte Carlo method.
Contrary to the linear
spin wave method the, the Monte Carlo method does not allow for
the determination of the density of states. Having the advantage
of being rotational invariant by construction, the Monte Carlo
method does not face the problem of the existence of zero energy modes.
We have discussed in detail what is the physics associated with
these modes. In the thermodynamic limit they play the role of
Goldstone modes, trying to restore the rotational symmetry of the problem,
explicitly broken by
the spin wave approximation. We have shown that in a numerical study
these modes can not be included in the calculation of operator
averages, if sensible physical results are to be obtained. This is because
these modes were already used in the construction of the broken symmetry
state, as was first discussed by P.~W.~Anderson in his seminal paper
on spin waves in non-diluted lattices.\cite{AND52}

Our approach allows us to compute
both the staggered magnetization and the N\'eel temperature as function
of the dilution concentration. In particular, the combination
of spin wave analysis and the recursion method allows for the
calculation of physical quantities virtually in the thermodynamic limit.
This possibility was not used before in similar studies on the square lattice.

We have used our results to explain the experimental data of two
Heisenberg honeycomb systems:  Mn$_{p}$Zn$_{1-p}$PS$_{3}$ (a diluted $S=5/2$
system)
and  Ba(Ni$_{p}$Mg$_{1-p}$)$_{2}$V$_{2}$O$_{8}$ (a diluted  $S=1$ system).
In the first case, the available experimental and theoretical studies
in the non-diluted regime suggest that second- and third-nearest-neighbor
interactions play a role on the physical properties of the system. This can
be seen from the fact that the measured magnetic moment of the samples
is finite beyond the classical site-dilution percolation threshold.
Our calculation suggests, however, that at low temperatures
and for $p>p_c$ the magnetic moment of these samples can be accounted for on
the basis of a single nearest-neighbor coupling. On the other hand, the
calculation of the N\'eel temperature using a  single nearest-neighbor coupling
is underestimated, as it should indeed be case based on the fact that
the magnetic order close to the N\'eel temperature should
have a measurable contribution from the other couplings, which are
not much smaller than the first nearest-neighbor coupling
(the N\'eel temperature for this system using second- and
third-nearest-neighbor interactions will be studied in a future publication).
Simple calculations based on simple (Ising like) mean
filed theories, on the other hand, are very much insensitive, by construction,
to the microscopic details of the system. Therefore, and as long as quantum
fluctuations are not important, a good agreement with the experimental
data should be obtained. This is the case for   Mn$_{p}$Zn$_{1-p}$PS$_{3}$,
but not for Ba(Ni$_{p}$Mg$_{1-p}$)$_{2}$V$_{2}$O$_{8}$ since its
much smaller spin brings about the contributions of quantum fluctuations.
In the case of the system  Ba(Ni$_{p}$Mg$_{1-p}$)$_{2}$V$_{2}$O$_{8}$,
there are, unfortunately, no measurement of its magnetic moment in the
diluted phase, however, the N\'eel temperature as function of dilution
is known from thermodynamic measurements. Our results show
that in this case, most likely, only the first-nearest-neighbor coupling
(and a very small magnetic anisotropy) are needed to describe the behavior
of the N\'eel temperature upon dilution. It would be important if further
investigations on this system could be performed in the future.

%%%%%%%%%%%%%%%%%%%%%%%%%%%%%%%%%%%%%%%%%%%%%%%%%%%%%%%%%%%%%%%%%%%%%%%%%%%%%%%
\subsection*{Acknowledgements}

Some of the understanding presented in this paper 
on the physics of the zero modes reflects a number of 
enlightening discussions with J. B. M. Lopes dos Santos, 
for which the authors are grateful.
We thank A. H. Castro Neto for illuminating conversations
of the physics of the 2D antiferromagnet in a square lattice. E.V.C.
acknowledges the Quantum Condensed Matter Theory Group at Boston University,
Boston, MA, U.S.A., for the hospitality, and the financial support of
Funda\c{c}\~ao para a Ci\^encia e a Tecnologia through Grant 
Ref. SFRH/BD/13182/2003. N.M.R.P. is thankful to the Quantum Condensed 
Matter visitors program at Boston University, Boston, MA, U.S.A., to the 
visitors program at the Max-Planck-Institut f\"ur Physik komplexer Systeme,
Dresden, Germany, and to Funda\c{c}\~ao para a Ci\^encia e a Tecnologia for 
a sabbatical grant. E.V.C., N.M.R.P. and J.L.B.L.S. were additionally financed
by FCT and EU through POCTI (QCAIII).

%%%%%%%%%%%%%%%%%%%%%%%%%%%%%%%%%%%%%%%%%%%%%%%%%%%%%%%%%%%%%%%%%%%%%%%%%%%%%%%
%%%%%%%%%%%%%%%%%%%%%%%%%%%%%%%%%%%%%%%%%%%%%%%%%%%%%%%%%%%%%%%%%%%%%%%%%%%%%%%
\appendix

%%%%%%%%%%%%%%%%%%%%%%%%%%%%%%%%%%%%%%%%%%%%%%%%%%%%%%%%%%%%%%%%%%%%%%%%%%%%%%%
\section{Diagonalization of $H_{\mathbf{k}=\mathbf{0}}$}
\label{sec:App-diagHk0}

The $\mathbf{k}=\mathbf{0}$ term in Hamiltonian \eqref{eq:Hk} can
be expressed as
\begin{equation}
H_{\mathbf{0}}=JSz\left(h_{a}(a_{\mathbf{0}}a_{\mathbf{0}}^{\dagger}
 +b_{\mathbf{0}}^{\dagger}b_{\mathbf{0}})+a_{\mathbf{0}}b_{\mathbf{0}}
 +b_{\mathbf{0}}^{\dagger}a_{\mathbf{0}}^{\dagger}\right).
\label{eq:Hk0}
\end{equation}
This is a standard bilinear model with two coupled modes, which is
straightforwardly diagonalized through a Bogoliubov-Valatin transformation
(Eq.~\eqref{eq:BVT}) when $h_{a}>1$. In the isotropic $h_{a}=1$
case it has an infinite number of eigenstates with a continuum energy
spectrum.

Let us define the following canonical transformation,
\begin{align}
a_{\mathbf{0}} & = \hat{q}_{1}+i\hat{p}_{1}\label{eq:a0q1p1}\,,\\
b_{\mathbf{0}} & = \hat{q}_{2}+i\hat{p}_{2}\,.\label{eq:b0q2p2}
\end{align}
We use the \emph{hat} notation to distinguish the operators from
their eigenvalues. The new generalized {}``position'' $\hat{q}$
and {}``momentum'' $\hat{p}$ operators satisfy the usual commutation
relations:
\begin{alignat}{3}
\bigl[a_{\mathbf{0}},a_{\mathbf{0}}^{\dagger}\bigr]&=1 & 
\quad &\Longrightarrow \quad &
\bigl[\hat{q}_{1},\hat{p}_{1}\bigr]&=\frac{i}{2}\,;\label{eq:q1p1cr}\\
\bigl[b_{\mathbf{0}},b_{\mathbf{0}}^{\dagger}\bigr]&=1 & 
\quad &\Longrightarrow \quad &
\bigl[\hat{q}_{2},\hat{p}_{2}\bigr]&=\frac{i}{2}\,.\label{eq:q2p2cr}
\end{alignat}
After simple algebra we find that the $h_{a}=1$ Hamiltonian \eqref{eq:Hk0}
can be written in terms of the new operators $\hat{q}$'s and $\hat{p}$'s as
\begin{equation}
H_{\mathbf{0}}=JSz\left[(\hat{q}_{1}+\hat{q}_{2})^{2}+(\hat{p}_{1}
 -\hat{p}_{2})^{2}\right].
\label{eq:H0qp}
\end{equation}
The variables $\hat{q}_{1}+\hat{q}_{2}$ and $\hat{p}_{1}-\hat{p}_{2}$
can be interpreted as the center of mass position and the relative
momentum, respectively, of a two particle system, therefore commuting
with each other
\begin{equation}
[\hat{q}_{1}+\hat{q}_{2},\,\hat{p}_{1}-\hat{p}_{2}]=\frac{i}{2}-\frac{i}{2}=0.
\label{eq:QPcr}
\end{equation}
Thus the eigenfunctions of Hamiltonian \eqref{eq:Hk0} are given
in as products of the eigenstates of the operator $\hat{q}_{1}+\hat{q}_{2}$
with eigenstates of the operator $\hat{p}_{1}-\hat{p}_{2}$,
\begin{equation}
\Psi_{Q,P}(q_{1},q_{2})=\delta(q_{1}+q_{2}-Q)e^{i\frac{P}{2}(q_{1}-q_{2})},
\label{eq:eigfuncHk0}
\end{equation}
and the aforementioned continuum spectrum is given by
\begin{equation}
E_{Q,P}=JSz(Q^{2}+P^{2}).
\label{eq:Eqp}
\end{equation}

%%%%%%%%%%%%%%%%%%%%%%%%%%%%%%%%%%%%%%%%%%%%%%%%%%%%%%%%%%%%%%%%%%%%%%%%%%%%%%%
\section{Anderson broken symmetry analysis}
\label{sec:App-Anderson}

In this appendix we closely follow the ideas developed by 
P.~W.~Anderson\cite{AND52} to show that the ground state of an 
antiferromagnet should display broken spin rotational symmetry, 
even in the absence of any anisotropy.

As was shown in Appendix \ref{sec:App-diagHk0} the operators 
$\hat{q}_{1}+\hat{q}_{2}$ and $\hat{p}_{1}-\hat{p}_{2}$ are constants of 
the motion in a system described by the isotropic Hamiltonian \eqref{eq:Hk0}, 
having well defined expectation values with zero dispersion. From definitions
\eqref{eq:a0q1p1} and \eqref{eq:b0q2p2}, and Eqs.~\eqref{eq:akbk}
and \eqref{eq:Sapprox}, it can be easily seen that
\begin{align}
\hat{q}_{1}+\hat{q}_{2} & = \frac{1}{\sqrt{2SN_{a}}}S_{\text{tot}}^{x}\,,
 \label{eq:QSx}\\
\hat{p}_{1}-\hat{p}_{2} & = \frac{1}{\sqrt{2SN_{a}}}S_{\text{tot}}^{y}\,, 
 \label{eq:PSy}
\end{align}
which explains the constant of motion character of the 
$\hat{q}_{1}+\hat{q}_{2}$ and $\hat{p}_{1}-\hat{p}_{2}$ operators 
($\mathbf{S}_{\text{tot}}$ is a constant of motion of the original 
Heisenberg model). The uncertainty relation ensures us that their 
canonical conjugates counterparts will have divergent dispersion. 
As for Eq.~\eqref{eq:QSx} and \eqref{eq:PSy} it is not difficult to 
show that the canonical conjugates of $\hat{p}_{1}-\hat{p}_{2}$
and $\hat{q}_{1}+\hat{q}_{2}$ are, respectively,
\begin{align}
\hat{q}_{1}-\hat{q}_{2} & = \frac{1}{\sqrt{2SN_{a}}}\biggl(\sum_{i\in A}
 S_{i}^{a,x}-\sum_{i\in B}S_{i}^{b,x}\biggr), \label{eq:q-mx}\\
\hat{p}_{1}+\hat{p}_{2} & = \frac{1}{\sqrt{2SN_{a}}}\biggl(\sum_{i\in A}
 S_{i}^{a,y}-\sum_{i\in B}S_{i}^{b,y}\biggr). \label{eq:p-my}
\end{align}

We want to know how much energy is needed to form a wave packet with
states \eqref{eq:eigfuncHk0} (above the ground state), such as the
expectation values of operators $\hat{q}_{1}-\hat{q}_{2}$ and 
$\hat{p}_{1}+\hat{p}_{2}$ have finite dispersion. Let us limit the 
fluctuations of the expectation value $\left\langle 
\hat{q}_{1}-\hat{q}_{2}\right\rangle $ to the
range $\Delta_{\hat{q}_{1}-\hat{q}_{2}}$. From the uncertainty relation
the expectation value of $\hat{p}_{1}-\hat{p}_{2}$ must now have
a nonzero dispersion, whose magnitude is given by
\begin{equation}
\Delta_{\hat{p}_{1}-\hat{p}_{2}}
 \approx\frac{1}{2\Delta_{\hat{q}_{1}-\hat{q}_{2}}}\,.
\label{eq:incert}
\end{equation}
Thus, to limit $\left\langle \hat{q}_{1}-\hat{q}_{2}\right\rangle $
to the range $\Delta_{\hat{q}_{1}-\hat{q}_{2}}$ we need
\begin{equation}
E_{lim}\simeq\frac{JSz}{4\Delta_{\hat{q}_{1}-\hat{q}_{2}}^{2}}\,,
\label{eq:Elim-mx}
\end{equation}
relatively to the ground state energy. Analogously, to limit 
$\left\langle \hat{p}_{1}+\hat{p}_{2}\right\rangle $ to the range 
$\Delta_{\hat{p}_{1}+\hat{p}_{2}}$ we need
\begin{equation}
E_{lim}\simeq\frac{JSz}{4\Delta_{\hat{p}_{1}+\hat{p}_{2}}^{2}}\,.
\label{eq:Elim-my}
\end{equation}
Defining the the mean square amplitudes of the quantities \eqref{eq:magx}
and \eqref{eq:magy},
\begin{align}
\sigma_{x}^{2} & = \frac{1}{(2N_{a})^{2}}\left\langle \left(\sum_{i\in A}
 S_{i}^{a,x}-\sum_{i\in B}S_{i}^{b,x}\right)^{\!\!2\,}\right\rangle\,,
\label{eq:VARmagx}\\
\sigma_{y}^{2} & = \frac{1}{(2N_{a})^{2}}\left\langle \left(\sum_{i\in A}
 S_{i}^{a,y}-\sum_{i\in B}S_{i}^{b,y}\right)^{\!\!2\,}\right\rangle\,,
\label{eq:VARmagy}
\end{align}
we find from \eqref{eq:q-mx} and \eqref{eq:p-my} that
\begin{align}
\Delta_{\hat{q}_{1}-\hat{q}_{2}}^{2} & = \frac{2N_{a}}{S}\sigma_{x}^{2}\,,
 \label{eq:VARq1-q2}\\
\Delta_{\hat{p}_{1}+\hat{p}_{2}}^{2} & = \frac{2N_{a}}{S}\sigma_{y}^{2}\,,
 \label{eq:VARp1+p2}
\end{align}
(note that $\hat{q}_{1}-\hat{q}_{2}$ and $\hat{p}_{1}+\hat{p}_{2}$
have zero expectation value). Inserting \eqref{eq:VARq1-q2} and 
\eqref{eq:VARp1+p2} in Eqs.~\eqref{eq:Elim-mx} and \eqref{eq:Elim-my} 
it can be seen that to limit $\sigma_{x}$ or $\sigma_{y}$ to a finite value we
only need an excess energy of magnitude $1/N_{a}$, and hence negligible
in the thermodynamic limit. As pointed out by Anderson, we can even
limit $\sigma_{x}$ or $\sigma_{y}$ to values of magnitude 
$1/N_{a}^{\frac{1}{2}+\alpha}$, with $\alpha>0$, requiring no energy when 
$N_{a}\rightarrow\infty$.

%%%%%%%%%%%%%%%%%%%%%%%%%%%%%%%%%%%%%%%%%%%%%%%%%%%%%%%%%%%%%%%%%%%%%%%%%%%%%%%
\bibliographystyle{apsrev}
\bibliography{./Bibliography/computationalphys,./Bibliography/condmattbooks,./Bibliography/magsitedilutexp,./Bibliography/magsitediluttheory,./Bibliography/recursionmethod}

\end{document}